\documentclass[12pt,epsf]{article}
\newlength{\abstractwidth}
\usepackage[dvips]{graphicx}
\usepackage{color}

\renewcommand{\thefootnote}{\fnsymbol{footnote}}
\renewcommand{\thanks}[1]{\footnote{#1}} % Use this for footnotes
\newcommand{\starttext}{
\setcounter{footnote}{0}
\renewcommand{\thefootnote}{\arabic{footnote}}}
\renewcommand{\theequation}{\thesection.\arabic{equation}}
\newcommand{\be}{\begin{equation}}
\newcommand{\bea}{\begin{eqnarray}}
\newcommand{\eea}{\end{eqnarray}}
\newcommand{\beq}{\begin{equation}}
\newcommand{\ee}{\end{equation}}
\newcommand{\eeq}{\end{equation}}

\def\<{\langle}

\newcommand{\half}{{1\over 2}}
\renewcommand{\>}{\rangle}
\def\bea{\begin{eqnarray}}
\def\eea{\end{eqnarray}}
\def\eq{&=&}

\def\f{phi}

\def\14{{1\over4}}
\def\12{{1 \over 2}}
\def\eq{&=&}

\def\d{\partial}

\def\h3{h^{3\over 2}}

\def\>{\rangle}
\def\<{\langle}

\def\d{\ensuremath{\Delta}}
\def\hyp{\ensuremath{{\cal H}_3}}

\def\f{\Phi}

\def\0cc{$\Lambda = 0$}

\def\wdw{Wheeler-DeWitt}
\def\sgg{$\Sigma$}
\def\rc{\ensuremath{{\cal R}}}
\def\tc{\ensuremath{{\cal T}}}

\catcode`\@=11
\@addtoreset{equation}{section}
\def\theequation{\arabic{section}.\arabic{equation}}
\catcode`@=12
\relax

\begin{document}
\renewcommand{\theequation}{\thesection.\arabic{equation}}
\begin{titlepage}
\bigskip
\rightline{SU-ITP-06-18} \rightline{OIQP-06-07}
\rightline{UCB-PTH-06/12} \rightline{LBNL-60518}
\rightline{hep-th/0606204}

\bigskip\bigskip\bigskip\bigskip

\centerline{\Large \bf {A Holographic Framework for Eternal Inflation }}

\bigskip\bigskip
\bigskip\bigskip
%\centerline{\it }
%\medskip
%\centerline{} \centerline{} \centerline{}
%\bigskip

\centerline{Ben Freivogel$^{a, b}$, Yasuhiro Sekino$^{c,d}$, Leonard
Susskind$^c$, Chen-Pin Yeh$^c$}
\medskip
\medskip
\centerline{\small $^a$Center for
  Theoretical Physics, Department of Physics, University of California, Berkeley}
\centerline{\small $^b$Lawrence Berkeley National Laboratory }
\centerline{\small $^c$Department of Physics, Stanford
University}
\centerline{\small $^d$Okayama Institute for Quantum Physics, 1-9-1
Kyoyama, Okayama 700-0015, Japan}
\medskip
\medskip

\bigskip\bigskip
\begin{abstract}
In this paper we  provide some circumstantial evidence for  a
holographic duality between  bubble nucleation in  an  eternally
inflating universe and  a Euclidean conformal field theory. The
 holographic correspondence (which is different than Strominger's
 dS/CFT duality) relates the decay of (3+1)-dimensional de Sitter space to
 a two-dimensional CFT. It  is not
 associated  with pure  de Sitter space, but  rather with  Coleman-De
 Luccia bubble nucleation. Alternatively, it  can be  thought  of as a
 holographic description of  the open, infinite, FRW cosmology that
 results from such a bubble.

The conjectured holographic representation  is  of a new type  that combines
holography with the  \wdw \ formalism to produce a \wdw \ theory that lives
on the spatial boundary of a $k=-1$ FRW  cosmology.

 We also argue  for a  more ambitious interpretation of the  \wdw \ CFT as a
 holographic  dual of the entire Landscape.
 \end{abstract}
\medskip
\medskip
\medskip
\bigskip
\bigskip
{\small {\bf Email:} freivogel@berkeley.edu, ysekino@v101.vaio.ne.jp,
susskind@stanford.edu, zenyeh@stanford.edu}

\end{titlepage}
\starttext \baselineskip=18pt \setcounter{footnote}{0}

%%%%%%%%%%%%%%%%%%%%%%%%%%%%%%%%%%%%%%%%%%%%%%%%%%%%%%%%%%%%%%%%%%%%%%
%%%%%%
%%%%%%%%%%%%%%%%
%%%%%%%%%%%%%%%%%%%%%%%%%%%%%%%%%%%%%%%%%%%%%%%%%%%%%%%%%%%%%%%%%%%%%%
%%%%%%
%%%%%%%%%%%%%%%%

\section{Introduction}
Over the  last decade two important  ideas--the Holographic Principle, and
the string theory Landscape--have radically changed our perspective, but the
consequences for  physics and  cosmology  are still largely unknown.

On the one hand the Holographic Principle~\cite{holography,
holography2} has  been   confirmed by Maldacena's discovery  of  the
AdS/CFT correspondence~\cite{adscft, adscft3}. There  is very little
doubt that bulk anti-de Sitter physics  is exactly dual to conformal
field theory on  the boundary of space. But we have no such
understanding of physics in an inflating (or accelerating)
background, i.e., our own world.

The second and more recent paradigm shift is  the  realization  that
string theory may have  a  vast Landscape of metastable de Sitter
vacua~\cite{landscape, landscape2}, and that the mechanisms of  eternal
inflation and bubble nucleation can populate  the  entire Landscape with
what Alan Guth  calls  ``pocket  universes"~\cite{eternal, guth}.
The Landscape  threatens to drastically alter our views of cosmology, and
also, the Laws of Physics themselves.

But one has the feeling  that the real  revolution is yet to come. It is hard
to believe that the two sets of ideas  are completely  unrelated. Indeed, the
truth is  that our understanding of de Sitter space and eternal inflation is
rudimentary at best, and until we know how to cast these  ideas  in a  rigorous
framework, the  entire structure is a very  rickety house  of cards.

 Our eventual goal  is  to bring together the two lines of thought--holography
 and Landscape--in  a comprehensive cosmological theory. We  will propose
 a holographic  dual description of the  multiverse  in the surprisingly simple
 form of a  two dimensional conformal  field theory. From one point of view,
 the more modest point  of view, the conformal field theory is dual to the
  Coleman-De Luccia (CDL) instanton description of false vacuum decay~\cite{cdl}.
  But we will argue for  a more expansive view in  which the CFT is dual to the
  entire multiverse. If true, it would follow that the Landscape would be encoded
  in a two dimensional conformal field theory.

 \subsection{A Conjectured Duality}
We will  consider  the  decay of a false vacuum with positive
cosmological  constant to a vacuum with vanishing vacuum energy, which presumably
lies on the flat moduli space  of  supersymmetric string vacua. However,  for
simplicity, we model the  Landscape by a simple scalar field potential  with two
minima. Later we will address  more  interesting realistic situations. We assume both vacua
are four dimensional; the extension to other dimensions is nontrivial.

\begin{figure}[!htb]
\center
\includegraphics [scale=.8]
{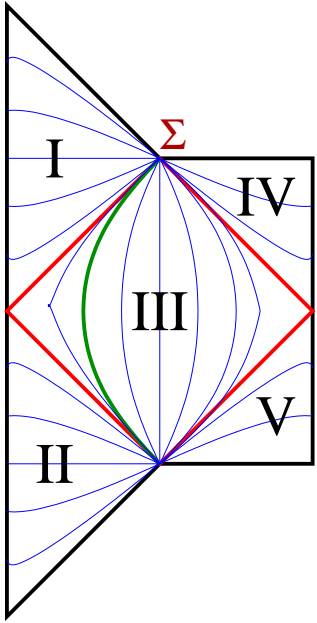} \caption{Penrose diagram
%for the decay of a de Sitter space to a flat space by CDL instanton.
for the Lorentzian continuation of the CDL instanton.
%Region~I and Region~II are asymptotically flat and Region~IV
%and Region~V are asymptotically de Sitter. Domain wall would be
%in Region~III.
Region~I is an open ($k=-1$) FRW universe which is asymptotically
flat. Region~IV is asymptotically de Sitter. $\Sigma$ is the
conformal 2-sphere defined by the
intersection of the light-like infinity of region~I and the
space-like infinity of region IV. The curves indicate orbits of the
$SO(3,1)$ symmetry, which acts on \sgg\ as the conformal group.}
\label{fig-penrose}
\end{figure}

The decay is described by the Coleman-De Luccia instanton. It is a
compact Euclidean solution with SO(4) symmetry which interpolates
between the true vacuum and the false vacuum. The geometry after
bubble nucleation is described by the continuation of the CDL
instanton to Minkowski signature,  the ``bounce geometry."  The
bounce geometry is a classical solution consisting of 5 regions
shown in Figure~\ref{fig-penrose}~\cite{cdl}.

Our focus will be on region I, which can be thought  of as the
interior of  the expanding bubble of  true vacuum. The metric in
region I takes the  form  of a standard infinite  open FRW universe
\be ds^2 = a(T)^2 (-dT^2 + dR^2  + \sinh^2R \  d\Omega_2^2).
\label{frw} \ee Under the  continuation, the $SO(4)$ symmetry
becomes the  non-compact symmetry $SO(3,1)$ which acts as  spatial
isometries on the constant time slices. These time slices are
uniform negatively curved hyperbolic geometries \hyp, isomorphic to
3-dimensional Euclidean Anti-de Sitter space.

The future boundary  of  the  geometry consists  of two portions.
One is the space-like  boundary of region IV, which is
asymptotically de Sitter.  The  other is  the light-like
``hat."\footnote{We focus on the future boundary because that is the
part of the geometry which appears after a bubble nucleation. In the
classical bounce geometry, Figure 1,
%which does not involve a bubble nucleation event,
the evolution from the past boundary up to the
``waist" of the diagram violates the second law, so it is
inconvenient to specify the ``in" state on the past boundary
\cite{boufre}. It is more natural to analyze the Hartle-Hawking
state on the future boundary, as we do here.} We will be  especially
interested in the intersection of these two which we call $\Sigma$.
$\Sigma$ is  an asymptotic 2-sphere that really
 represents spatial infinity in the FRW patch.

The central point about $\Sigma$ is that the $SO(3,1)$ symmetry  group acts on it as
2-dimensional conformal transformations. Each time slice is geometrically equivalent
to a Euclidean AdS space, and has exactly the same symmetries.
This suggests that $\Sigma$ may be the location of a holographic dual description.

The two-sphere $\Sigma$ is really spacelike infinity when viewed from the
  interior of the open FRW geometry  of region I. From the Holographic Principle,
  it follows that there must exist a boundary description of this  region, and it
  should have its degrees of freedom on $\Sigma$. We have seen that  $SO(3,1)$  acts
  as conformal symmetry  on  $\Sigma$, and that  naturally raises the  question
  of whether field correlators extrapolated to $\Sigma$ might be holographically
   dual to a (Euclidean) conformal field  theory (CFT)  in a manner  analogous to
   the familiar AdS/CFT duality. We might hope that such  a CFT is the
    holographic description of  the  open  FRW universe in region I.

  If so, there would be  a surprising twist. Instead of a  $2+1$ dimensional hologram,
  $\Sigma$ would be a $2$ dimensional Euclidean hologram. Not only would the space-like
  ``depth" direction   have to emerge from the CFT, but  time would also be emergent.
  Indeed we will find evidence for such a mechanism in the form of a Liouville field in the boundary theory.

As we will see in the rest of the paper, the fact that the
Euclidean geometry of the CDL instanton (a deformed sphere) is
compact has a profound consequence for holography. We review the geometry in the next section. Although the FRW region
of the Lorentzian geometry has infinite volume, the compactness
of the Euclidean space seems to forbid a quantization
with a fixed boundary condition, in contrast to
the case of AdS space where both the Euclidean and Lorentzian
geometries are non-compact.

This will become clear when we study the correlation functions for
massless fluctuations in section \ref{sec-corr}. The correlators in
the FRW region which are obtained by  analytic continuation from
Euclidean space do not decay at large spatial separation. This
implies that a non-normalizable mode, which is usually considered to
be fixed by the boundary conditions and therefore non-dynamical, is
now path-integrated.\footnote{Our results differ from the existing
literature, e.g. \cite{openinflation, sasaki, turok1}, which claims
that the non-normalizable contribution is pure gauge. }

Our computation of the graviton correlator is the main evidence that
the dual theory is a local 2-dimensional CFT. We find a piece of the
graviton correlator which is transverse, traceless, and dimension 2
in the boundary theory. This piece has the right properties to be
dual to the energy-momentum tensor of the CFT.

The presence of a non-normalizable graviton mode means that the
geometry of the boundary fluctuates. In the dual CFT, we expect that
the fluctuations of the boundary geometry are described by a
Liouville field, as we explain in section \ref{sec-holcor}. Because
of the fluctuating boundary, the bulk theory is  similar in some
ways to quantum gravity in a closed space. The general formalism to
treat such a system is the Wheeler-DeWitt theory, which we review in
section \ref{sec-wdw}. Since the metric is a dynamical variable, we
cannot treat time as a parameter. Instead, time must be defined in
terms of clocks which are part of the dynamical system. A convenient
choice of clock that is often used is the scale factor, which
ordinarily is monotonically increasing for open FRW universes.

What emerges from our considerations is a new kind of hybrid--a
holographic WDW theory--in which bulk information is holographically
encoded in boundary degrees of freedom, but in which time emerges as
a dynamical degree of freedom as in WDW theory. The Liouville field
will play the role of time as does the scale factor in the bulk WDW
theory. To allow such an interpretation, the Liouville field should
have negative metric, which is well known to be the case in the 2-D
gravity coupled to a large number of matter fields. The existence of
negative metric fields in the CFT is suggested from the bulk
correlator calculated in section~\ref{sec-corr}; it implies that the
CFT is not unitary.
 We will discuss properties of the
holographic WDW theory in section~\ref{sec-holwdw}.

In section \ref{sec-inst} we point out that a true vacuum bubble
inevitably collides with an infinite number of other bubbles
\cite{guthweinberg}. If the de Sitter space can decay to more than
one true vacuum, bubble collisions have a dramatic effect on the
boundary \sgg, creating regions which are in a different vacuum. We
interpret these as instantons in the dual CFT. We see this as a
suggestion that the matter sector of the CFT may contain the entire
Landscape as its target space.
%Since in our more ambitious
%interpretation the CFT would be dual to the entire string theory landscape,
%which includes vacua which are not four-dimensional, we need the CFT to
%contain instantons whose interior is effectively higher dimensional.

We describe a number of open questions in section \ref{sec-open}.
The details of our calculation of the correlation function are in
Appendix A for the case of the scalar field and Appendix B for the
graviton.

We work in four bulk dimensions throughout. This is not without
consequences. Our preliminary calculations suggest that in any
number of bulk dimensions the boundary geometry will fluctuate. But
only in a 2-dimensional boundary theory can we describe fluctuations
of the geometry in terms of a Liouville field; in higher dimensions
the situation is unclear.

\subsection{Related Work}

The duality that we are proposing is different from the ``dS/CFT
correspondence,'' proposed by Strominger~\cite{dscft}.
%(See also~\cite{maldacenadscft})
%in the sense that our duality is motivated by a belief that
%the bulk and boundary theory has the same Hilbert space
%and Hamiltonian.
In dS/CFT, the dual CFT is assumed to be at the 3-dimensional
spatial surface at the future infinity of de Sitter space.
Although de Sitter correlators have conformally invariant form,
the late time behavior should be affected by the nucleation of
infinitely many bubbles, and the basis for the duality is not
entirely clear~\cite{dyson}. If it exists and makes sense, it
describes a global view of cosmology that cuts across causally
disconnected regions separated by event horizons. The theory in
this paper is more local and describes a single pocket universe
that emerges from the local decay of a metastable de Sitter
vacuum.

Holographic duals containing gravity have been proposed in various
contexts similar but different from ours. It has been noted that if
the AdS space is cut off at finite volume, the dual theory should
contain gravity, essentially because the gravity on the boundary
becomes normalizable and fluctuates~\cite{verlinde, adsgravity}.
Brane-world models where two cut-off AdS (or Schwarzschild AdS)
spaces are patched at the cut-off boundary have been studied from
the perspective of holographic duality. The holographic dual theory on
the brane contains gravity. Fairly concrete analyses have been done
in~\cite{adsgravity, hms}.

In~\cite{dsds}, an intersting proposal was made for an approximate
holographic description for a causal patch of de Sitter space. On
the basis of an observation that the limit of de Sitter geometry
near an observer's horizon approaches the limit of AdS geomerty far
from the boundary, the authors proposed a CFT dual for de Sitter
valid in the low energy limit. It is argued that the CFT contains
gravity since the geometry where the CFT is expected to live can
fluctuate as in the brane-world cases.

There have also been attempts to describe de Sitter space or
cosmology using the AdS/CFT correspondence by embedding the spacetime
in asymptotically AdS space. Ref.~\cite{shenker} pursues the
possibility of extracting information on the bubble of de Sitter
space embedded in an asymptotic AdS space from the boundary CFT.
Ref~\cite{horowitz} studies a model of big crunch in AdS/CFT
correspondence.

Our proposed duality has similarities with the proposal of de Boer
and Solodukhin \cite{deboer} for a holographic theory of flat space.
They considered the hyperbolic slicing of flat space (a part of it
corresponding to our region I), and proposed that a dual theory
should be a CFT living on the boundary of the hyperboloid; this is
precisely what we do. However, we benefit from having a compact
Euclidean geometry. In contrast to~\cite{deboer} where they consider
a continuous infinite family of the CFT operators to reconstruct the
flat space correlators, we will naturally find a discrete tower of
operators in the CFT. In addition, our main focus is on the piece of
the correlation function which is absent in flat space.

 \section{The CDL Instanton}

In this section we review vacuum decay via Coleman-de Luccia tunneling, focusing on the simplest case. We imagine a potential which has a false vacuum with positive vacuum energy and a true vacuum with zero vacuum energy.

 The scalar field that  interpolates between the two  vacua, and its
 potential,  will be called $\f$ and $V(\f)$. The two minima are at
 field values $\f =\f_F$ and  $\f=\f_T$:
 \bea
 V(\f_F) &>& 0 \cr
 V(\f_T) \eq 0.
 \label{V}
 \eea
The Euclidean signature CDL instanton geometry has the  topology of a 4-sphere.
It is described by the  metric
\be
ds^2 = dt^2  + a(t)^2 d\Omega_3^2
\label{cdl}
\ee
where $d\Omega_3^2$ is the  metric of  a unit 3-sphere.
The Euclidean equations of motion are
\bea
\left({\dot{a} \over a}\right)^2  \eq \ \half \dot{\f}^2 -V(\f) +
 {1\over a^{2}} \cr
\ddot{\f} \eq -3{\dot{a} \over a} \dot{\f} + {\partial V \over {\partial \f}}.
\label{eom}
\eea
The Euclidean time runs from $t=0$ to $t=t_0$
and the boundary conditions are
\bea
\dot{a} \eq1, \ \ \ \  \ (t=0) \cr
\dot{a} \eq -1,\ \ \ (t=t_0)\cr
\dot{\f} \eq 0, \ \ \ \ \ (t=0, \ \ t=t_0).
\label{boundcond}
\eea
The most important features of the  instanton geometry are its topology, $S^4$,
and  its symmetry, $SO(4)$.  Note  that  in general the $SO(5)$ symmetry of $S^4$
is broken. The $SO(4)$ acts on the 3-sphere $\Omega_3$.

It will be convenient to  change variables from $t$ to a
``conformal" variable $X$ defined  by \be dX = {dt \over a(t)} ~.
\label{conf} \ee In terms of $X$, the metric becomes \be ds^2 =
a(X)^2(dX^2 + d\Omega_3^2)~. \ee The variable $X$ runs  over the
entire real axis and the equations of motion become \bea (a')^2 \eq
\half a^2 (\Phi')^2 -a^4 V(\Phi) +a^2 \cr \Phi'' +{2a'\Phi' \over a}
\eq a^2 {\partial V \over \partial \Phi}, \label{xeqs} \eea where
prime indicates derivative with respect  to $X$.

 As described in  the introduction, the continuation to Minkowski signature defines the bounce geometry, which is relevant after the tunneling event.
 In   order to continue the Euclidean CDL geometry  to region
I of the Minkowski geometry (see figure \ref{fig-penrose}), we write the 3-sphere metric in the  form
\be
 d\Omega_3^2=d\theta^2 +\sin^2{\theta} d\Omega_2^2.
 \label{threesphere}
\ee The continuation is defined  by \bea X &\to& T +  {\pi\over 2}i
\cr \theta &\to& iR. \label{cont} \eea The metric in region I is an
open FRW universe, \be ds^2 = a(T)^2 (-dT^2 + dR^2  + \sinh^2R \
d\Omega_2^2) = a(T)^2(-dT^2 + d\hyp ^2) \ee The $SO(3,1)$ symmetry
is realized on the hyperboloid \hyp. \hyp\ is equivalently a
3-dimensional Euclidean AdS space. Defining \sgg\ to be the boundary
of \hyp, it is clear that the symmetries act on \sgg\ as the
conformal group, just as in the AdS/CFT correspondence.

  \section{Correlation Functions}
  \label{sec-corr}

The basis  for  our  conjecture is the properties of correlation
functions of ordinary bulk  fields in  the  bounce geometry. The
method of calculation  is to   begin with correlation functions
defined on the original Euclidean CDL geometry. Analytically
continuing the Euclidean correlation functions  to Minkowski
signature then allows us to extrapolate  them to $\Sigma$. One can
then ask if the correlation functions have the properties required
of fields in a CFT. The method of calculation is explained  in
detail in Appendix A for scalars, and in Appendix B for gravitons.
In this section we will discuss the general features. Since we
compute the correlators by analytic continuation from Euclidean
space, what we are doing is computing expectation values in the
Hartle-Hawking state~\cite{hh}. Because we are computing expectation
values of Hermitian operators with arguments at space-like separated
points, our answers are guaranteed to be real.
%This fits well with the dual CFT, which we expect to be a
%Euclidean theory containing operators with real conformal
%dimensions.
This fits well with the dual CFT, which we expect to be a Euclidean
theory. As we will see, the bulk correlators can be interpreted as
correlators of CFT operators with real dimensions.

We examine the correlators in a particular limit. We believe that the CFT
 should be closely related to the late time asymptotics of the correlators,
 where particles become non-interacting; we think of the CFT as describing the
  ``out" state. So we want to take the limit $T \to \infty$. Furthermore, as usual
   it is useful to think of the CFT as living on the boundary $R \to \infty$ of AdS
   space, in this case \sgg, so we want to expand the correlators for large $R$. To be
    precise, we look at the correlators in the limit
\be
T + R  \to \infty
\ee
and in an expansion which is valid for large $R - T$.

  \subsection{Massless Scalar}
We will begin  with the  simplest case of a minimally coupled  massless bulk scalar
 field $\chi$, then move to the case of the graviton. We can make use of the symmetries
 to write the correlator as a function of a few variables. Recall that the metric in region I is
\be
ds^2 = a(T)^2(-dT^2 + d\hyp^2).
\ee
The $SO(3,1)$  symmetry means that the correlator cannot depend
arbitrarily on the location of the two points in the hyperboloid \hyp;
it must be a function of only one parameter, which we take to be the
geodesic distance $\ell$ on the unit hyperboloid. The correlator can
depend on the two time coordinates in an arbitrary way because we have
no time translation invariance.

We are primarily interested in the piece of the correlator which has
non-trivial dependence on $\bar{T}=T_1+T_2$. This piece does not
exists in the flat space; it essentially corresponds to the particle
production due to the geometry.

Calculation of the Euclidean correlator boils down to a
one-dimensional Schr\"{o}dinger problem, and the $\bar{T}$ dependent
piece is written in terms of the reflection coefficient $\rc (k)$.
In Appendix A we derive an expression for the Lorentzian correlator
in the form of a contour integral,
\begin{equation}
\langle \chi(T_1,0)\chi(T_2,R)\rangle^{(\bar{T})}
=e^{-\bar{T}}\oint_C {dk\over 2\pi} \rc (k)e^{-ik\bar{T}}{
\left(e^{-ikR}-e^{ikR-2\pi k}\right) \over 2\sinh R \sinh k\pi},
\label{maincontourint}
\end{equation}
where we have placed one of the points at the origin of the
hyperboloid, so that the geodesic distance is $\ell=R$.
The last factor comes from the Green's function on $S^3$, which we
rotate to \hyp.

\begin{figure}[!htb]
\center
        \includegraphics[angle = 90,width = 4.5 in,height = 2 in]{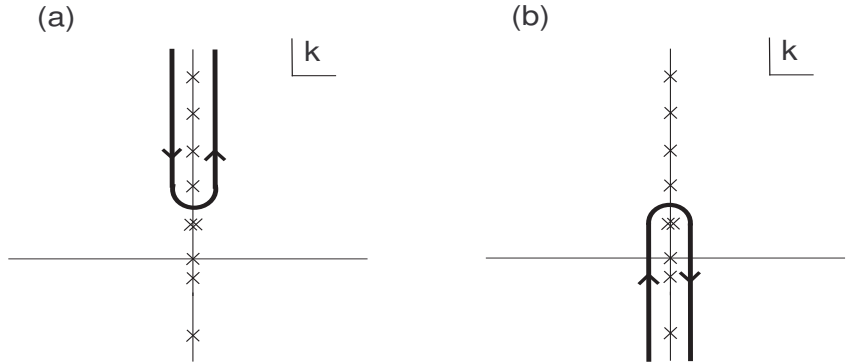}
        \caption{(a) Contour surrounding single poles
at $k=2i,3i,\ldots$. (b) Coutour surrounding a double pole at $k=i$,
single poles at $k=0,-2i,-3i,\ldots$, and a single pole at $k=-ia$
$(0<a<1)$ which is due to the pole of $\rc (k)$; for the
``half-sphere'' example in Appendix A.2, $a=1/2$. There is no pole
at $k=-i$ since $\rc (k)$ has a zero there.} \label{fig-Lorentz}
\end{figure}

In the limit of interest, the two terms in the integrand are
evaluated on different contours. The first term is closed in the
lower half plane, as in Figure~\ref{fig-Lorentz}(b), while the
second term is closed in the upper half plane,  as in
Figure~\ref{fig-Lorentz}(a). The reflection coefficient has a pole
at $k=i$, as does $1/\sinh (k\pi)$, so there is a double pole at
$k=i$. This double pole is ultimately due to the zero mode in
Euclidean space. The factor $1/\sinh (k\pi)$ gives rise to single
poles at integer multiple of $k=i$, and the correlator is written as
a sum of an infinite number of terms.
%The regularly spaced single poles coming from
%$1/\sinh(k \pi)$ can be thought of as the discrete momenta of the
%spherical harmonics on $S^3$.
% which we rotated to become the hyperboloid \hyp.

The correlator consists of two
types of terms
\be
  \langle \chi \chi \rangle = G_1(T_1, T_2, \ell)+G_2(T_1, T_2, \ell),
\ee
as we explain in the following.

\subsubsection{Terms with definite dimensions}
The first term
$G_1(T_1, T_2, \ell)$ arises from the contour shown in figure \ref{fig-Lorentz}(a), together with one piece of the double pole contained in the contour shown in figure \ref{fig-Lorentz}(b). It has the form
 \be G_1 = \sum_{\d =2}^{\infty} G_\d
~e^{(\d -2)(T_1+T_2) }~ {e^{-(\d -1)\ell} \over \sinh \ell}
\label{geeone} \ee where the $G_\d$ are constants. The motivation
for the unusual label \d\ for the summation variable will become
clear in a moment. This formula has an interesting interpretation.
Each term in the sum is the product of a time-dependent piece and
a piece which depends on the separation of the points on the
hyperboloid $\hyp$. The latter, \be {e^{-(\d -1)\ell} \over \sinh
\ell}~, \label{cob} \ee
 is exactly the correlation function for a  (fictitious) 3-dimensional
 AdS bulk field of mass
 \be
 m^2 = \d (\d - 2)
 \ee
  in Euclidean AdS. So the correlator appears to consist of an infinite series of terms,
  each corresponding to a field with definite mass in Euclidean AdS.

As usual, the mass of a field in AdS translates into the conformal dimension in the dual CFT.
To see the relationship, note that
 the geodesic distance $\ell$ can be expressed in terms of $R_1, R_2, \Omega_1, \Omega_2$.
 In  the limit that the $R$'s tend to infinity (the  limit in  which the points tend  to $\Sigma$),
 $\ell$ has  the form
 \be
 \ell \to R_1 +R_2  + \log(1-\cos \alpha),
 \ee
 where $\alpha$ is  the angle between the points  on the 2-sphere.
The  correlator (\ref{geeone}) can be rewritten in this limit
\be
e^{(\d-2)(T_1+T_2)}{e^{-(\d -1)\ell} \over \sinh \ell} \to
e^{(\d-2)(T_1+T_2)}
e^{-\d R_1} e^{-\d R_2}(1 - \cos{\alpha})^{-\d}.
\label{foureight}
\ee
Apart from the  time dependent factors, these expressions are familiar from the
AdS/CFT duality where they represent correlation functions of boundary conformal
 fields of dimension $\d$. The factors $e^{-\d R}$ are wave function renormalization
 factors and the angular dependence $(1 - \cos{\alpha})^{-\d}$ is the renormalized two
 point function of a field of dimension \d. The obvious interpretation  is that the
  boundary theory has an infinite number of conformal fields of dimension $\d= (2,3,4....)$.
  In particular the lowest dimension field has dimension $\d=2$. Note that the dimension
  of the Lagrangian of a 2-D CFT is 2.

Defining the light-cone variables
\bea
T^+ \eq T+R \cr
T^- \eq T-R,
\eea
and combining (\ref{geeone}) and (\ref{foureight}) we can write the correlator in the form
\be
e^{-(T^+_1 +T^+_2)} \sum_{\d} G_{\d} e^{(\d -1)(T^-_1 +T^-_2)} (1-\cos{\alpha})^{-\d}.
\label{fourten}
\ee

The characteristic behavior of this term is simple: it  has  an overall
factor $e^{-T^+}$ for each external ``leg." If we strip off that factor the  remaining
 expression has a smooth  limit as we approach the hat at $T^+ = \infty$. The remaining factor
 is  a sum of discrete contributions  that can be identified with  correlators of definite
 dimension $\d$.

Equation (\ref{fourten})  suggests a strategy for connecting the CFT
correlators to the correlators %not on \sgg \ but rather
on the hat.
These correlators are roughly analogous to S-matrix elements
involving  massless outgoing particles.
%The form of (\ref{fourten})
%suggests first of all, that the asymptotic behavior of bulk fields
%behaves like $\exp{(-T^+)}$. Stripping off this factor would leave
%finite correlators on the hat.}
Equation (\ref{fourten}) suggests that each bulk scalar field be
identified with an infinite  sum of boundary conformal fields, \be
\lim_{T^+ \to \infty} \  e^{T^+} \chi = \sum_{\d} e^{(\d-1) T^-}
\chi_{\d}. \ee

\subsubsection{Logarithmic terms}
The second term $G_2$ arises from the double pole at $k=i$. It is
more difficult to interpret. It has the form \be G_2=  {e^\ell
\over \sinh \ell} (T_1 +T_2 +\ell), \label{second} \ee which for
large $R_1,R_2$ behaves like \be G_2 \sim  (T_1 +T_2 + R_1 +R_2 +
\log(1-\cos \alpha)). \label{geetwo} \ee

We will interpret this as follows. The minimally coupled massless
scalar has a translation symmetry $\chi \to \chi+\chi_0$ where
$\chi_0$ is a constant. Interpreting this as a gauge
 symmetry, the gauge invariant quantities are derivatives of $\chi$. Thus we are really
 interested in derivatives of the correlator of the form
\be
\partial_{1} \partial_{2} G_{2}.
\ee It is  evident from (\ref{geetwo}) that only the last term
contributes when we differentiate with respect to both arguments of
$G$. Thus for gauge invariant purposes we may  replace
(\ref{geetwo})
 by the simpler form
\be
G_2=  \log(1-\cos \alpha).
\label{geewhiz}
\ee
The expression in (\ref{geewhiz}) has a familiar significance. $G_2$ is
exactly the correlation function of a massless dimension-zero scalar
field in the boundary theory.

However, from the bulk point of view, something unusual is happening.
The behavior (\ref{geetwo}) indicates that
the fluctuation of $\chi$ contains \it non-normalizable \rm modes
on the hyperbolic space \hyp. In AdS space such modes are associated
with boundary conditions and not dynamical degrees of freedom.
What we are finding is that the infrared behavior of the correlation
function  describes not only  normalizable quantum fluctuations in
the background FRW universe, but  also  fluctuations of the  asymptotic
boundary conditions. In other words the infrared behavior of the correlation
contains information about  a measure on the space of asymptotically
different pocket universes.

From the form of (\ref{second}) we see that  there  is a sub-leading term.   Writing
$$
{e^{\ell} \over \sinh \ell} = 2 + {e^{-\ell} \over \sinh \ell},
$$
we see that the sub-leading term in  the  limit $R \to \infty$ behaves like
\be
\left[T_1 + T_2 +R_1 +R_2 + \log(1-\cos \alpha)\right] e^{-2\ell}
\label{ugh}
\ee
 Apart from the factor $[(T_1 + T_2 +R_1 +R_2 + \log(1-\cos \alpha)]$ this  is  the
  correlator for a dimension 2 field. Evidently, this  term is corrupting the dimension
  2 piece of the  correlator with  a prefactor that tends  to infinity. If we make the
  usual connection between the  radial  variable $R$ and scale-size in the boundary theory,
   then the prefactor would represent a logarithmic running  of the  correlation function.

\subsection{Separating the Double Pole}

In fact the logarithmic behavior associated with (\ref{second}) is
probably not present in the correlator of a realistic scalar field.
It is associated with the infrared divergences of a massless
minimally coupled (MMC) bulk field in the compact Euclidean CDL
geometry--divergences due to the zero mode of the field. Now MMC
scalars may  exist in the  final vacuum with zero cosmological
constant. Indeed the only such vacua in string theory are on the
supersymmetric moduli space and the moduli themselves are MMC
scalars. But one expects that in a stable de Sitter vacuum the
moduli will be fixed. Thus one should allow the scalar field to have
a mass in the false vacuum. The effect of adding such a mass is to
split the double pole at $k=i$ into a pair of single poles at $k=i$
and at $k=i (1- \epsilon)$. The result is the following (more
details in Appendix A.4):

1) For small mass the logarithmic term (\ref{second}) is replaced by
a term with positive dimension equal to $\epsilon$. As the mass
increases the dimension also increases until it approaches dimension
$\epsilon =1$ at some threshold mass.
%Beyond that mass, the pole at $k= i(1- \epsilon)$ disappears.
Beyond that mass, the pole moves to the lower half plane and ceases
to contribute to the correlator in our limit.

2) The corrupted dimension $2$ term splits into two contributions,
each with definite dimension. One term coming from the pole at $k=i$
has dimension $2$. The other term moves to $\d = 2+\epsilon$.

Our conclusion from this latter result is that there are two
distinct terms with $\d = 2$ and $\d = 2+\epsilon$ which
``accidentally" become degenerate as the mass term is switched off.

%Although the  infrared behavior is improved,
%the dimension still violates the  Breitenlohner-Freedman bound and still
%describes non-nomalizable fluctuations.

\subsection{Gravitational Correlators}

The description of bulk gravitational fluctuations $h_{\mu}^{~\nu}$
in the transverse traceless gauge is very similar to the case of the
MMC scalar. Much of the analysis of that case carries over. In
Appendix~B we calculate the two  point  function $\langle
h_{\mu}^{~\nu}  h_{\mu'}^{\ \nu'}   \rangle$. Since the  symmetry of
the problem acts on \hyp, we decompose the graviton in
representations on \hyp; our interest is in modes which are
transverse and traceless on \hyp. Here we quote the results from
Appendix~B.

As in  the scalar case,  the correlator contains two terms, one
essentially  identical to (\ref{geeone}) except that the constants
$G_\d$ are replaced by  bi-tensors describing the index structure.  As
in the scalar case the leading dimension is 2.
We find a dimension 2 piece which is transverse-traceless
not only in the three-dimensional  bulk space but also in the two
dimensional boundary sense.
%The tensor structure of the dimension 2 term is transverse traceless
%not only in the three-dimensional  bulk space but also in the two
%dimensional boundary sense.
In other words if we define the indices on  the asymptotic 2-sphere $\Sigma$
to be $i,j$, then this dimension 2 piece satisfies
%\bea
%\nabla_i G^{i  j}_{\ k l} \eq 0  \cr
%G^{i  j}_{\ i l} \eq 0
%\label{tt}
%\eea
\bea \nabla_i \langle h^i_{~j} h^k_{~l}\rangle_{(2)} \eq 0  \cr
\langle h^i_{~i}  h^k_{~l}\rangle_{(2)} \eq 0. \label{tt} \eea
Apart from a numerical factor of proportionality, this piece is a
candidate for the energy-momentum tensor of the boundary theory.
The equations (\ref{tt})  become conservation and tracelessness of
the stress tensor.

As in  the scalar case we find a term in the correlation function
which is similar to (\ref{geetwo}).
The correct expression is  given in equations (\ref{gravitylog}) and
(\ref{tensort}).
Once again, to get a gauge invariant quantity, we need to take
derivatives of the correlator.
%Once again, only derivatives of $h_\mu^{~\nu}$ are gauge invariant.
%An example  is  the fluctuation of the curvature invariant of the
%asymptotic  2 sphere
Let us consider the fluctuation of the curvature invariant
of the asymptotic 2-sphere
\be
C(\Omega_2) = \partial_i \partial^j h_j^i.
\ee
Explicit calculation shows that the correlation function of  the  two
dimensional  curvature satisfies
\be
\langle C C\rangle = (1-\cos{\alpha})^{-2}
\label{CCC}
\ee
so that the curvature is an operator of dimension 2.

The curvature fluctuations of the boundary sphere  \it are  of the same
order of magnitude as the background curvature.  \rm   The meaning  of
this is that the shape of  the asymptotic 2-sphere--its oblateness for
example--has finite fluctuations even  as $R$ and $T$ go to
infinity. Note that by contrast, in AdS space, the fluctuations of the
shape of the boundary go to zero for all finite energy
configurations. Another way to say the same thing is that in AdS the
dynamical quantum fluctuations of the space are  composed of
normalizable modes
%while in  the present case they are not normalizable.
while in the present case non-normalizable modes
also fluctuate.
%
%Explicit calculation shows that the correlation function of
%the  two  dimensional  curvature satisfies
%\be
%\langle C C\rangle = (1-\cos{\alpha})^{-2}
%\label{CCC}
%\ee
%so that the curvature is an operator of dimension 2.

The fluctuations of the intrinsic geometry of   the boundary
$\Sigma$ require  us to replace
 $d\Omega_2^2$ in (\ref{frw}),  with
\be e^{L(\theta,\alpha)}d\Omega_2^2 \label{liou}
\ee
where $L$ is a
Liouville  field~\cite{liouville} on $\Sigma$. The existence of a
Liouville field in the boundary theory clears up a mystery.
Ordinarily, holographic theories are $D-1$ dimensional  with one  of
the dimensions being time. But in  the  present case we are
proposing a $D-2$ dimensional Euclidean holographic  dual. We have
lost a dimension but gained a  Liouville  field.

Now let's  return to the dimension $2$ piece of the correlator. As
in the scalar case there are logarithmic terms that corrupt the pure
$\d =2 $ term, once again from a double pole at $k=i$. If we split
the double pole into single poles, then as in the scalar case, we
find two operators. The one at $\d =2$ is transverse-traceless in
the boundary directions but the shifted one at $\d= 2+ \epsilon$ is
not. It is
%the term from the single pole at $k=i$
the former that we identify with the stress tensor of the boundary
theory.

The logarithmic factor for the dimension 2 piece is a result of the
degeneracy of the $\d=2$ piece and $\d= 2+\epsilon$ piece which have
opposite signs. The term with negative sign in the correlator
implies the presence of a field with negative metric in the CFT.
This is what we want. We expect the Liouville field to be a negative
metric field since we want to interpret it as time; the Liouville
field can have negative kinetic term when there are large number of
matter fields. The existence of a field with negative kinetic term
suggests that the CFT is non-unitary.

\section{The Holographic Correspondence}
\label{sec-holcor}

In the usual AdS/CFT duality the extra bulk direction of space
corresponds to scale size in the  CFT~\cite{edlen, uvir}.
 In Poincare coordinates the  spatial metric is
\be
ds^2 = a^2{dz^2 + dx^idx^i \over z^2}
\label{ads}
\ee
where $a$ is the overall radius of the AdS space, $x^i$ are the coordinates in the plane of the boundary, and $z$ is  the  emergent radial coordinate.

It is well known that the image of  a point located at $z$ is a patch on the boundary of coordinate size
\be
\Delta x =z.
\ee

Thus  we see that scale size  in the CFT is dual to the  emergent coordinate $z$. Equivalently we can write
\be
\Delta x = e^{-R}.
\ee
Thus the coordinate $R$ is logarithmically related to scale size  in the boundary  theory
\be
R = -\log{\d x}
\ee

Now let us  regulate the boundary sphere by locating  it  at finite $R$. More exactly, let the boundary  be at a variable radial distance given by
\be
R=R(\Omega_2).
\ee
This  is  not enough to define  the regulated sphere because all the  constant $T$ surfaces intersect on \sgg.

Let us compare the situation with AdS/CFT. In that case the emergent
direction of  space would correspond to $R$, the spatial direction
perpendicular to $\Sigma$. In fact $e^{-R}$ is identified with scale
size in  the CFT.  The geometric arguments for  that identification
should go over into  the present case. But now  consider the fact
that $\Sigma$ is really a place where all time slices intersect.
Specifying a large sphere involves not only choosing a large value
of $R$ but also a value of FRW time. In fact we can introduce a
local time at each point of the sphere, $T(\Omega_2)$, or
$T(\theta,\alpha)$. From (\ref{frw}) and the  fact that \be a(T)
\sim e^T, \ee
  we see that the  metric of the  asymptotic 2-sphere is  proportional  to
 \be
 e^{2T(\theta,\alpha)} d\Omega_2^2.
 \label{tliou}
 \ee
 Comparing (\ref{liou}) and (\ref{tliou}) we see that there is  a natural identification of the Liouville field
 with local time,
 \be
 L(\theta,\alpha)= 2T(\theta,\alpha).
 \ee

\section{Wheeler-DeWitt Theory}
\label{sec-wdw}
 As in the case of anti-de Sitter space, the spatial slices of open FRW are negatively curved hyperbolic planes with a boundary at spatial infinity. In the  present  case that  boundary is the  infinite 2-sphere $\Sigma$.  It is entirely natural to expect that a holographic description  of open FRW lives on this  boundary. However there is one big difference between open FRW and AdS. It is not that the  background is time dependent in the FRW case. Time dependence can easily be introduced into the AdS/CFT correspondence, for example by varying the value of the gauge coupling constant in a time dependent way. More important is that in the AdS case the boundary conditions at spatial infinity are frozen. To vary the boundary conditions requires exciting the non-normalizable modes which always costs infinite energy. By contrast, the boundary conditions in open FRW are not frozen.

 The fact that spatial infinity is not frozen has implications for the  way we think  about time. Unlike AdS we cannot  define a global time variable by introducing synchronized clocks at infinity. It follows that a precise notion of global energy such as ADM energy does not exist.  Not even a time dependent global Hamiltonian makes sense. What option do we have  to define a quantum mechanics of  open FRW?

 The answer in  our opinion  is to construct a new kind of hybrid of the
 Wheeler-DeWitt theory~\cite{wdw} and the Holographic Principle.  Let us begin with
 a quick review  of the Wheeler-DeWitt theory.
One begins with the Einstein equations. Define
 \be
E^{\mu}_{~\nu} = R^{\mu}_{~\nu} -\12 \delta^{\mu}_{~\nu}  R  -\kappa T^{\mu}_{~\nu}.
 \ee
 Four of the  equations are considered to  be  constraints that act on
 the \wdw \ wave function
 \bea
 E_{~a}^0 \Psi(g_{ab},f)\eq 0 \cr
  E_{~0}^0 \Psi(g_{ab},f)\eq 0.
  \eea
The wave function $\Psi$ is  a functional  of the space-space components
of the  metric $g_{ab}$ and the matter fields which we summarize by $f$.
The mixed space-time components are somewhat trivial and merely state  that the wave function is  invariant under spatial reparameterizations. The time-time equation is the interesting  one that contains  all  the dynamics. It can be written  as
 \be
 H \Psi(g_{ab},f)= 0
 \ee
 where $H$ is   the Hamiltonian density  derived from the  gravitational  and matter action.

Typically there are many solutions to the \wdw \ equations corresponding to different boundary conditions but the interesting point is that  the  wave functions are independent of time. They depend only on spatial  geometry and matter fields. Any sense of time and temporal evolution must be emergent.

There are many ways to extract time from the \wdw \ wave function by  identifying some dynamical variable  as a clock. The most common is to use the FRW scale  factor which in conventional cosmology is  a monotonic function of time.  The logic is nicely  illustrated in  the  so called mini-superspace  approximation in which all the fields  are taken to  be spatially homogeneous and  the geometry is assumed isotropic. In  that  case the spatial  geometry  is  characterized by  a scale  factor $a$.

To derive the \wdw \ equation we begin with the Lagrangian
%(all inessential constants set equal to $1$)
 \be {\cal{L}}= v \left\{ [-a
 \dot a^2  +ka ] + [a^3{\dot f^2 \over 2} -a^3 V(f)] \right\} \ .
 \ee
The constant $k $ is equal to $1$, $0$, or $-1$ for  closed, flat,
or open spatial geometries, and $v$ is the coordinate volume of
space; we are setting all inessential constants equal to $1$, so $v$
is dimensionless. The two terms in square brackets represent the
gravitational and matter field action. From this  we deduce the
Hamiltonian \be H= -{p_a^2 \over 4av} +{p_f^2 \over 2a^3 v} + v a^3
V(f) - vka \ee
 where $p_a$ and $p_f$ are the  momenta conjugate to $a$ and $f$. If we make the  identifications
\bea
p_a \eq  -i \partial_a  \cr
p_f \eq  -i \partial_f,
\eea
 then the  \wdw \ equation becomes
 \be
 \left({\partial_a^2 \over 4av} -{\partial_f^2 \over 2a^3v} +va^3 V(f) -vka  \right) \Psi (a,f)=0.
 \label{msswdw}
 \ee

 When supplemented by various boundary conditions the  \wdw \ equation  determines the  time  independent wave function $\Psi(a,f)$. But the full interpretation  of the  equation requires one  additional ingredient that follows from the Lagrangian--namely
 \be
 \dot a ={p_a\over 2av}.
 \label{adot}
 \ee
 Equation (\ref{adot}) provides the connection between time and scale factor that allows us  to  think  of the scale factor  as a  clock. The physics is  illuminated by first solving (\ref{msswdw}) setting the matter energy to  zero. We will first illustrate the method in the flat case $k=0$ (a torus geometry for example) and with the  matter energy being replaced by a cosmological  constant
 \be
 \left({\partial_a^2 \over 4av} + va^3 \lambda  \right) \Psi (a)=0.
 \ee
 There are two independent  solutions corresponding to expanding ($\dot a >0$) and contracting  ($\dot a <0$) universes. The expanding solution, for large $a$ is given by
 \be
 \Psi= e^{ica^3v}.
 \label{Psi}
 \ee
 where the constant $c$ is  given  by $c = 2 \lambda^{\12}/3$.
%In the semi-classical limit, the role of time is played by
%$\log a$; the relation (\ref{adot}) implies $\dot{a}=3ca/2$,
%on the state (\ref{Psi}).

 Now if  we  reinstate the matter field and write
 \be
  \Psi= e^{ica^3v} \psi(a, f)
 \ee
 we find that for large $a$ (\ref{msswdw}) reduces to the conventional
Schr\"{o}dinger  equation
 \be
 -i\partial_t \psi(a,f)  = H_{matter}\psi.
 \ee
with the  role of time being played by $\log{a}$; in the semi-classical
limit, the relation (\ref{adot}) implies $\dot{a}=3ca/2$, which is
given by evaluating (\ref{adot}) on the state (\ref{Psi}).

% Note that in (\ref{Psi}), the exponent in $\Psi$ is proportional to  the  volume of  space $a^3 v$.

Beyond the mini-superspace context the single clock represented by the scale factor is replaced by
local clocks. For example, the  determinant of the space-space metric  defines a local scale factor by
 \be
 |g| \to a(x)^6.
 \ee
 The local scale factor provides a local clock at each  point  of space. We may also define
 a global clock by  averaging the local clock over space.

Note that in (\ref{Psi}), the exponent in $\Psi$ is proportional to
the  volume of  space $a^3 v$. We can obtain this wave function
alternatively by dividing space into cells with volume $v_{\rm
\scriptsize cell}$, and using the mini-superspace approximation for
each cell. The leading term in the wave function at large $a$ will
be the direct product of the wave functions for single cells
 \be
 \Psi = \exp ({\sum_{\rm \scriptsize cells}i c  a^3 v_{\rm \scriptsize cell}} )\ .
 \ee
The exponent is now proportional to the sum of the volumes of each cell, so again it is
proportional to the total volume of space.
%In particular, it is independent of how we divide the space up into cells.

% We can go slightly beyond the
% minisuperspace approximation by dividing space into a number of cells and using the
% approximation in each one. The leading term in the wave function at large $a$ will be the
% generalization of (\ref{Psi}),
% \be
% \Psi = \exp ({\Sigma_{\rm \scriptsize cells}i c  a^3 v_{\rm \scriptsize cell}} )\ .
% \ee

In the open negatively curved FRW universe without cosmological
constant, there is a qualitative difference.
% It is especially interesting to apply  the mini-superspace method to
% the open negatively  curved FRW universe without cosmological
% constant.
%
% The proper way to do this is to divide or tessellate the
% hyperbolic space into cells, each one about the size of the background
% radius  of curvature. The mini-superspace approximation can be thought of as applying to a single cell.
This time the WDW equation for large $a$ is
 \be
 \left({\partial_a^2 \over 4av} + va  \right) \Psi (a)=0,
 \ee
and the wave function for an expanding  universe has  the form
 \be
 \Psi \sim e^{2ia^2v}.
 \label{openwdw}
 \ee

%Again, we can go beyond the minisuperspace approximation by dividing
%space into cells.
Again, let us divide the space into cells to illustrate the point.
For simplicity, let us assume that the cells have the coordinate
size equal to the curvature radius of the hyperboloid. The leading
behavior of the wave function would be the direct product
 \be \Psi \sim \exp({\sum_{{\rm \scriptsize cells}}2i
 %v_{{\rm\scriptsize cell}}
 a^2}) \ .
 \label{openbig}
 \ee

% The sum appearing in the exponent of (\ref{openbig}) is  odd. Each term contains the
% coordinate volume of the cell multiplied by $a^2$. This is unnatural because $a^2$
% would naturally multiply an area rather than a volume. However, recall that in hyperbolic
% space the area is proportional to the volume as long as the region in question is big
% compared to the radius of curvature. Presumably our approximation is only good in this
% regime, so we may replace the coordinate volume by the coordinate area. Now each term
% in the sum is the area (measured in Planck units) of the particular cell that it refers
% to.
%
% However, now we have another confusion. Adding up areas over the  volume of space
%  is an unusual operation. Adding up volumes gives the total  volume but adding up areas
%  has no sensible meaning--except in  hyperbolic space. To see the point, first regulate
%  the space by giving  it a boundary at some large value of $R$. The point about
%  hyperbolic space is that almost all of the cells are  near the boundary. The sum is
%  dominated by cells within the radius of curvature of the boundary. Adding up  their area
%  simply gives the area of the  entire regulated region. Thus the WDW wave function has a
%  distinctly holographic flavor. For late time (large $a$) it has the form
 The sum appearing in the exponent of (\ref{openbig}) is  odd. Each
 term is the area of the particular cell (in Planck unit).
 Adding up areas over the  volume of space is an unusual operation.
 Adding up volumes gives the total  volume but adding up areas
 has no sensible meaning--except in  hyperbolic space.
To see the point, first regulate the space by giving  it a boundary
at some large radius $R$. The point about hyperbolic space is that
almost all of the cells are  near the boundary. The sum is dominated
by cells within the radius of curvature of the boundary. Adding up
their area simply gives the area of the regulated boundary. Thus the
WDW wave function has a distinctly holographic flavor. For late time
(large $a$) it has the form
 \be
 \Psi \sim e^{i \alpha  ({\rm \scriptsize Area})}\ ,
 \ee
where $\alpha$ is a number of order one and the area is measured in
Planck units.
%We have seen that this answer is independent of how we
%divide space into cells as long as the cells are big enough.

 For further details on the  \wdw \ formulation and the  emergence of
 time  we refer the  reader to \cite{fischler} and especially
 \cite{banks}.

 \section{Holographic Wheeler-DeWitt Theory}
\label{sec-holwdw}

 According  to the  Holographic Principle the degrees of freedom describing a region of space can be identified with the region's boundary. In the  familiar cases like AdS, the boundary is frozen in the sense that it takes an infinite energy to excite the non-normalizable modes that have  support at the  boundary.
 The conceptual importance of this fact is  that one  can always imagine adding distant  physical clocks to the system without perturbing the interesting physics deep in  the interior. These asymptotic clocks provide the definition of time that forms the  basis for  the ADM formalism in which the  Hamiltonian is  a surface integral. In AdS, it is the ADM Hamiltonian which becomes identified with the Hamiltonian of the dual holographic boundary theory.

The present case of open  FRW is different. As we have  seen the  boundary geometry is not frozen. The  only way to introduce asymptotic clocks is to build them  out of  the  asymptotic degrees of freedom  already present in the  system. An obvious candidate  for a local boundary clock is the local scale factor, i.e., the Liouville  field. We therefore  propose the following scheme.

Assume that the boundary \sgg \ is  equipped with a set of fundamental holographic degrees of freedom whose precise nature is unknown. We  would expect that  these degrees of  freedom would transform nontrivially under some type of  gauge transformations and that only  gauge invariant combinations would  correspond  to  limits  of bulk fields.  Among the  degrees of freedom, or perhaps  composed  out of them, is a Liouville field $L(\Omega_2)$. For simplicity we label the fundamental holographic fields as ($L, f$).

We also assume that there  is  a  local generator of time translations on \sgg \ that has  the form  of  a  Hamiltonian density $\hat{H}(\Omega_2)$ and a  \wdw \ wave function $\hat{\Psi} (L,f)$ satisfying
\be
\hat{H}\hat{\Psi}=0.
\ee
The quantity  $\hat{\Psi}^{\dag} \hat{\Psi}$ defines the  measure for  computing  expectation  values of functions of  $L$ and $f$. With no loss of generality we  can write
\be
\hat{\Psi}^{\dag} \hat{\Psi} = \exp\left\{-S(L,f)\right\}.
\label{S}
\ee
 In other  words the functional $S(L,f)$ defines  an action for  a  2-dimensional Euclidean CFT on  \sgg. However what  we don't know  is whether  $S$ has  the form of a local action, i.e., an integral of  a local density.

 \subsection{Locality?}
 Is there reason to hope that the holographic WDW wave function
 might be local in the above  sense? It is difficult to prove or
 disprove, but the following
argument might be suggestive.

Let us  consider a regulated  version  of  the  theory with a boundary
placed at a finite  distance $R$. Let $\alpha$ be an angle on the sphere
and let us measure time by  the conformal time $T$.

From the form of the open FRW metric one sees that the coordinate
speed of light on the boundary sphere tends to zero like $e^{-R}$.
%The implication is  that the spatial derivative terms in the
%boundary Hamiltonian $\hat{H}$ must also tend to zero.
We see this as a suggestion that the spatial derivative terms in the
boundary Hamiltonian $\hat{H}$ must also tend to zero.
One could locally undo this  slowing down by re-scaling the space
coordinate on the  sphere by  a compensating factor.

Now  (for finite $R$) we write the  WDW wave function in  the form (\ref{S}).
Cluster decomposition requires $S$ to have the form of an integral over $\Sigma$ of
connected clusters (monomials) where the coefficient functions in the clusters tend
to  zero  at large separation.
But as $R$ tends  to infinity the coordinate size of the  clusters shrinks due
to the re-scaling of the velocity of light.  This  suggests that the wave function
may become local.

On the other hand, the final correlation functions that we would obtain from the wave
function are not local. They have short distance singularities but they are not
delta functions or derivatives of delta
functions. There is no  contradiction since a local action can lead to non-local
correlations. However, the naive argument about the  locality of $\hat{\Psi}$ can be
criticized on the grounds that if applied  directly to the correlation functions, it
would  say that  they are local.

One interesting point is that the leading term in the bulk
wave function is local. As we saw in the last section,
the wave function has  the leading behavior $\exp\{ic ({\rm  Area})\}$,
and the area is the  ultralocal integral over \sgg
\be
{\rm Area} \ \ = \int_{\Sigma} e^L.
\ee

A strict test of locality would be the existence of a local
boundary energy-momentum tensor. We have found a candidate whose
2-point function satisfies the crucial conditions of being
traceless and transverse, and has the correct operator dimension.
However this is far from sufficient.
Calculating three-point functions could confirm that operator
product expansions involving the boundary stress tensor satisfy
the usual conditions.

In what follows we  will  assume the locality of $S$ without proof.

 \subsection{Properties of the CFT}
The first question about a CFT is  ``What is its central  charge?"
This can be read off from the two point function of the
energy-momentum tensor but to do so, we need the numerical factor
connecting the energy-momentum tensor with the metric fluctuations.

Let  us  assume that  the bulk metric fluctuation is canonically
normalized.
 In  that  case it has the same units  as a  canonically  normalized scalar. In
 four space-time dimensions the  field has units of inverse length. Consider the
 value of the two point function on the  Euclidean CDL instanton geometry.
 If we assume the critical bubble is of the same scale as the Hubble radius in the de Sitter space,
 then the only
 scale in the problem is the Hubble radius of the false vacuum. We suspect that even if the critical bubble is parametrically small compared to the Hubble radius, our results continue to hold.
 Let us call the Hubble radius $D$.
 Evidently the  two point function must scale like
 \be
 \langle hh\rangle \sim D^{-2}.
 \ee
 Now  let us  assume that the  2-dimensional $T_{~\mu}^{\nu}$ is  proportional  to
 the  boundary value of
 $h$. Schematically
 \be
 T = q h
 \ee
 with $q$ a numerical  constant. It follows that the short distance singularity
 in $\langle  TT\rangle$ will have strength
 \be
 \langle TT\rangle \sim q^2/D^2.
 \ee

 Next consider the  3-point  function $\langle hhh\rangle$. One thing we know
 about  it is that  it contains  a gravitational coupling  constant $\kappa$
 with units of length. We also  know that  it  has dimensions of inverse  length
 cubed. Thus it must  be of order
 \be
 \langle hhh\rangle \sim \kappa/D^4
 \ee
 and
 \be
  \langle TTT\rangle \sim q^3\kappa/D^4.
 \ee

 Finally, we use the  fact that the  ratio of  the  two  and three point
 function is controlled by the classical algebra of diffeomorphisms.
 Schematically $[T,T]=T$. This  implies that the ratio of the two and three
 point function is parametrically independent of $D$ and $\kappa$. Thus
 \be
 q \sim D^2/\kappa
 \ee
 and the  two  point function is of order
 \be
 \langle TT \rangle \sim D^2/\kappa^2 = D^2 /G
 \ee
 The implication is that the central charge is of order the entropy of
the parent de Sitter space that gave birth to the  open FRW by CDL
tunneling.

In a Liouville theory the central charge is not a measure of the
number of degrees of freedom of the system. One can see that in the
semiclassical limit in which the total central charge can be written
as a sum of a Liouville term $c_L$, and a matter term $c_M$.
Typically (and we will assume it here) the matter central charge is
positive and represents the number of matter degrees of freedom. But
the Liouville central charge can be negative, especially when the
Liouville field represents a time-like coordinate.

Assuming that the Liouville theory can be decomposed into a 2-D gravitational
sector and a 2-D matter sector, one can ask if the  matter sector is conformal.
The coupling of Liouville to a conformal field  theory has especially simple
properties that make it easy to study.
However any such theory has an extra Weyl invariance that can be interpreted as
unbroken translation invariance of the  Liouville field.
That would imply time-translation ($T$ translation) invariance which is
not something we expect in a cosmological theory.

One direct way to see that the matter sector is not conformal is
from the curvature-curvature correlator (\ref{CCC}). When Liouville
is coupled to conformal matter, the two sectors decouple and the
Liouville field satisfies a field equation that implies no
fluctuation in the curvature. When coupled to non-conformal matter,
the curvature (more precisely, the Euler density, $\sqrt{g}R$)
becomes a fluctuating field with dimension $2$. Evidently
(\ref{CCC}) implies that the matter sector in not
conformal\footnote{The curvature correlator (\ref{CCC}) has a
special property that it is independent of $T$ (the Liouville
field). This fact also needs an interpretation in the CFT.}.

It is clear that there is a kind of Infrared/Ultraviolet connection
in the system. The Liouville degree of freedom can be identified
with the logarithm of the ultraviolet regulator scale.
%\be
%L = \log{\kappa}
%\ee
Since the Liouville field is identified with time, it is evident
that the late time behavior of the FRW universe is controlled by the
UV behavior of the CFT.

What is the UV behavior of the matter theory?  The Holographic Principle
suggests an answer. The maximum entropy in a bulk region of fixed
radius--say $R$-- is proportional to the proper area of the boundary
\be
S_{max} \sim e^{2R} e^{2T}.
\label{smax}
\ee

On the boundary theory side the number of degrees of freedom can be estimated
in a manner analogous to the method used in \cite{edlen} \ but with some modifications
required by the Liouville field. As in \cite{edlen},
one first regulates the
theory by introducing a coordinate length scale $\delta$ on the boundary sphere.
The cutoff $\delta$ is not a proper-distance cutoff. It represents a uniform
grid size on
the background unit 2-sphere.  If there were no Liouville field,
 the number of effective degrees of freedom in the cutoff area
 is of order $c_m$. In this counting, a single 2-dimensional Dirac fermion
 is a single degree of freedom.

However, the existence of a Liouville field modifies this formula so that the
number of degrees of freedom is
\be
%\delta N_{dof} \sim c_m \delta^{2} e^L.
N_{dof} \sim c_m \delta^{-2} e^L. \label{ndof} \ee If the matter
theory is not conformal, then $c_m$ should be thought of as a scale
dependent quantity.

In \cite{edlen} it was explained that an ultraviolet cutoff in the
CFT is equivalent to an infrared cutoff on the radial coordinate
$R$. The cutoffs are related by \be \delta = e^{-R}. \ee Evidently,
matching  the number of degrees of freedom of the Liouville theory
with the number expected from the Holographic principle, requires
$c_m$ to tend to a constant in the ultraviolet. This means that
 the matter theory is governed by a  fixed point in the UV.
%An interesting speculation is that the UV value of the matter central charge
%may depend on dynamics.
%
%The UV value of the matter central charge will depend on dynamics.
%For example, if a period of slow roll inflation occurs inside the
%true vacuum bubble, then we expect the amount of matter entropy
%within one curvature radius to increase exponentially with the
%number of e-foldings. This suggests that $c_m$ may depend strongly
%on the amount of inflation.

At the moment that is all we know about the matter theory but we hope to come
back to it.

\section{Boundary Instantons and Bubble  Collisions}
\label{sec-inst}

We have  assumed that the  world can be described by a single CDL bubble.
However, Guth and Weinberg
long ago  showed that the  isolated pocket universes that are produced by eternal
inflation are in fact infinite clusters of colliding bubbles~\cite{guthweinberg}.
In this  section we will discuss the implications of such  collisions
for the holographic boundary description.

\begin{figure}[!htb]
\center
\includegraphics [scale=.4]
{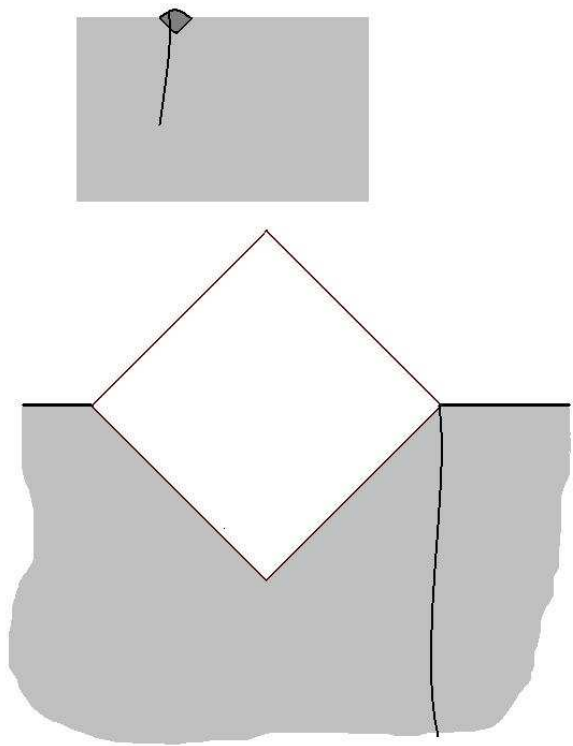} \caption{Top: Time-like trajectory that ends up in a
bubble. Bottom: Time-like trajectory that meets a flat-space bubble
at $\Sigma$.
%intersects the future boundary at $\Sigma$.
Since infinitely many bubbles, which are not drawn here, form in de
Sitter space, this trajectory will inevitably be caught in another
bubble like the one in the top figure.} \label{fig-CFTpix}
\end{figure}

To understand why bubbles must occur in infinite clusters consider a
time-like trajectory in  pure de Sitter space that ends at the future
boundary. Assume that there is a finite probability per unit  proper
time for a bubble to nucleate and engulf the trajectory. Then, since the
trajectory is infinite, it is  certain that a bubble will nucleate
before the trajectory reaches the future boundary. This is  shown in
Figure \ref{fig-CFTpix} (top figure).

%FIG
%
%Now consider the time-like trajectory in fig
%
%FIG
%
Now consider the time-like trajectory that intersects the
future boundary  at $\Sigma$ (bottom of Figure \ref{fig-CFTpix}).
The proper  time of such a trajectory is infinite so that it is
certain that a bubble nucleates along it. Such a bubble will definitely
collide with the original bubble. When  bubbles collide they
in effect form a single  bubble but the
intersection of the bubble boundary with the future
boundary of de Sitter space (we still call that intersection $\Sigma$)
 will no longer be a geometric sphere.
 This is  another indication that the  geometry  of  $\Sigma$ fluctuates.
 These fluctuations are clearly non-perturbative since bubble
 nucleation  is a  tunneling  process.

In order to understand the importance of such  processes let us  consider a
Landscape  that includes a
key  feature, namely a moduli space of vacua with
vanishing vacuum energy.
In string theory this would be the  supersymmetric moduli space (SMS).
In general it is possible to decay to different points on the moduli space.
In fact we  doubt that  there are any obstructions to decaying  to
arbitrary points on the  SMS. This means that a given de Sitter vacuum can decay
to a variety  of bubble-types.

Consider a decay to a point R on the moduli space. Let's call it the Red point.
At a later time (global de Sitter time) a
 bubble nucleates with the interior at the Blue point B. Assume the two points
 are connected by massless moduli. The boundary sphere (topologically spherical)
 $\Sigma$ consists of two regions, one red and one blue. The blue region
 will be very much smaller than the red if the blue decay is later than the red.
 On the other hand, in  the bulk, the  red and blue will bleed into one another
 and as time evolves the bulk space should evolve to some average color. But the
 boundary $\Sigma$ remains divided into red and blue.

\begin{figure}[!htb]
\center
\includegraphics [trim=1cm 4cm 3cm 3cm, clip, scale=.4]
{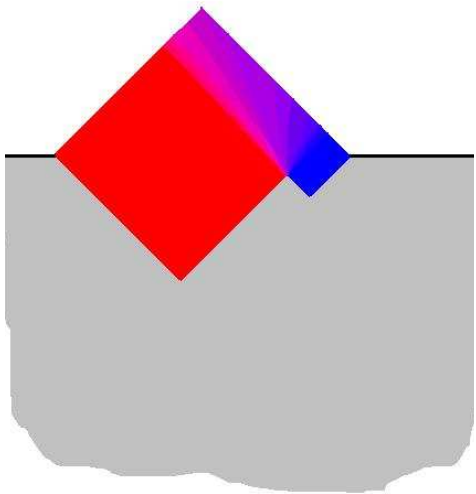} \caption{Collision of two flat-space bubbles of different
types.}
%The field value in the bulk will be smoothed out, but the
%boundary will remain divided into two domains.}
\label{fig-blood}
\end{figure}

The blue patch in the red background has a natural interpretation in
the boundary field theory. It is a  localized non-vacuum region with
an exponentially suppressed amplitude. In other words it is an \it
instanton. \rm

The statement that the blue  vacuum nucleates later than  the  red is
not invariant. There are symmetries of the  background de
Sitter space  that interchange the  order of events. This manifests itself
in the boundary  theory. A conformal transformation of the sphere $\Sigma$
can push the blue region out into the red region and shrink the red to a
small patch. Turning an instanton ``inside out" is always  possible in a
conformal field theory.

Another interesting point: a red bubble can nucleate and collide with the
blue bubble. This will appear as a red instanton inside the blue instanton.

\begin{figure}[!htb]
\center
\includegraphics[trim=0cm 0cm 0cm 0cm, clip, scale=.3]
{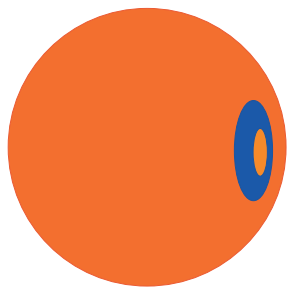}
%\caption{An instanton inside another instanton.}
\caption{A red instanton inside a blue instanton.}
\label{fig-blueinstanton}
\end{figure}

The integration over instanton sizes  is equivalent to the integration
over the time at which the associated bubble nucleates. Since one expects
the rate of nucleation to be constant, the integration must diverge. The
divergence is for small instanton sizes from the CFT point of view.
The  implication is clear: the average number of instantons must be infinite and be
dominated by instantons of  arbitrarily small size.

%The instanton gas reflects the diversity of the Landscape.

\begin{figure}[!htb]
\center
\includegraphics[trim=1cm 3cm 3cm 6cm, clip, scale=.45]
{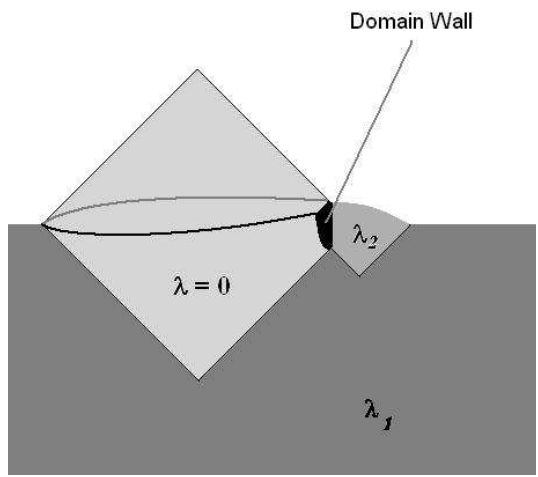} \caption{Collision of two bubbles with different
cosmological constants.} \label{fig-THREELAMBDAS}
\end{figure}

So far we have considered instantons associated with decays onto the
moduli space. Let us consider another situation in which the
original bubble collides with a bubble of non-zero vacuum energy
$\lambda_2$ as in Figure~\ref{fig-THREELAMBDAS}. The two bubbles,
one with vanishing cosmological constant, and one with positive
cosmological constant will be separated by a domain wall, and the
surface \sgg \
 will have two portions adjacent to $\lambda_1$ and $\lambda_2$.

It appears that instanton gas reflects the  diversity of the
Landscape. This suggests that the matter sector of the Liouville
theory is rich enough to contain, as part of its target space, the
entire Landscape including the supersymmetric moduli space as well
as all the vacua of non-zero cosmological constant.

We began by considering a transition from a particular de Sitter
vacuum to a point on the space of vacua with vanishing cosmological
constant. The tentative result is a Liouville theory with central
charge determined by the entropy of the false de Sitter vacuum. But
then it seems that non-perturbative effects will drive us to a
democratic view of the space of vacua; the Liouville theory is
associated with the entire Landscape and not just two points. The
simplified model of just two points may only be an approximation of
a very small part of the theory.

If there is a master Liouville theory of the whole Landscape, then
what is its total central charge? We suspect the answer may be zero.
Perhaps it is determined by the configuration of maximum vacuum
energy, i.e., the Planck energy. A de Sitter space with a Planckian
cosmological constant would have vanishing entropy.  But obviously
this is taking us into very deep waters so we will quit here before
we drown.

\section{Open Questions}
\label{sec-open}

There are a number of questions about our proposal that we left
unanswered.

First and foremost is whether the WDW wave function really defines
a local conformal field theory.  Equation (\ref{S}) defines an
action but we don't know if it is  local. An obvious test is to
show that the candidate energy-momentum tensor satisfies the usual
properties. Thus far we have calculated the two-point function and
found it to be traceless, transverse, and of dimension $2$. That
is as far as we can go without calculating three-point functions.
With three point functions one can test various operator product
expansions and also the Schwinger or Virasoro algebra required of
a local field theory.

Another question that needs study is what is the effect of vacua
with negative cosmological constant. Nucleation of  such bubbles leads to
singular crunches. What happens when a singular bubble
collides with a bubble of zero cosmological constant? Does such a collision
give rise to additional instantons?

%Since in our more ambitious interpretation the CFT would be dual to
%the entire string theory landscape, which includes vacua which are
%not four-dimensional, we need the CFT to contain instantons whose
%interior is effectively higher dimensional.

In our more ambitious interpretation, the CFT would be dual to the
entire string theory Landscape. The Landscape includes vacua which
are not four-dimensional. What will happen when the four-dimensional
universe meets a region with different dimensionality? Does the CFT
contain instantons whose interior is effectively higher dimensional?

Finally, what are the degrees of freedom of
the  boundary theory? Do they derive  from some
kind of gauge theory? And how does  it all fit together with string theory?

%%%%%%%%%%%%%%%%%%%%%%%%%%%%%%%%%%%%%%%%%%%%%%%%%%%%%%%%%%%%%%%%%%%%%%%%%%%%%%%%%%%%%%%%%%%%%%%%%%%%%%%%%%%%%%%%%%%%%%%%%%%%%%%%%%%%%%%%%%%%%%%%%%%%%%%%%%%%%%%%%%%%%%%%%%%%%%%%%%%%%%%%%%%%%%%%%%%%%%%%%%%%%%%%%%%%%%%%%%%%%%%%%%%%%%%%%%%%%%%%%%%%%%%%%%%%%%%

\subsection*{Acknowledgements}
We would like to thank Raphael Bousso, Eric Gimon, Alan Guth, Simeon Hellerman,  Juan Maldacena,  Alex Maloney, John McGreevy, and especially
Steve Shenker for discussions.

\section*{Appendix}

\appendix

%%%%%%%%%%%%%%%%%%%%%%%%%%%%%%%%%%%%%%%%%%%%%%%%%%%
%%%%%%%%%%%%%%%%%%%%%%%%%%%%%%%%%%%%%%%%%%%%%%%%%%%
%%%%%%%%%%%%%%%%%%%%%%%%%%%%%%%%%%%%%%%%%%%%%%%%%%%

\section{Scalar correlator in the CDL geometry}
In this appendix, we calculate the two-point functions of a
scalar field in a 4-dimensional open FRW universe generated
by the Coleman-De Luccia (CDL) instanton.
The cosmological constant in the true vacuum is assumed to be zero.

We study the correlator of a massless minimally coupled
scalar in detail.
The main reason for studying this is its similarity
with the graviton correlator.
%%  At the end of this appendix,
%% we will also study a scalar which
%% is massless in the flat part of the geometry and
%% is massive in the de Sitter part of the geometry.
%This should be qualitatively similar to the actual moduli
%fields.
Our method for calculating the correlator is to first
obtain the correlator on the Euclidean CDL instanton
background and then analytically continue the expression
to the FRW region (region I of figure \ref{fig-penrose}) of the Lorentzian geometry.
This corresponds to taking the Hartle-Hawking vacuum in the bounce solution.
Analysis of the scalar correlator in CDL geometry using
this method has been done by Gratton and Turok~ \cite{turok}.
(See also refs.~\cite{openinflation} for studies of the perturbations
the CDL background.)

We will often use the thin-wall approximation.
In the thin-wall limit, the Euclidean geometry is a flat disc
patched to a portion of a sphere.
In this limit, the FRW region of the Lorentzian geometry
is exactly a piece of Minkowski space, but the non-trivial geometry in the other
regions affects the correlation function.
The thin-wall approximation is helpful since we can
find the massless scalar correlator exactly. We will
derive asymptotic behaviors of the correlators using
the thin-wall approximation, but we will argue that the
conclusion is true for general CDL geometries.
%study asymptotic behavior of the correlator in the
%large distance limit in the Lorentzian geometry.
%Our discussion is based on thin-wall case, but we argue
%that the conclusion applies to the general CDL geometry.

%Main reason for studying the massless minimally coupled
%scalar is that it is similar to the graviton.

In section A.1, we explain our prescription
for calculating massless scalar correlators. Then in A.2,
we present an explicit calculation in a particular
thin-wall geometry. We obtain an exact expression for the correlator
in A.2.1, and its large distance behavior is
studied in A.2.2. In A.2.3, we give an alternative way
to derive the asymptotic behavior.
%In fact, the purpose of A.2.1 is to
%illustrate the procedure for getting the correlator,
%and the reader does not have to follow all the details.
%There is a simpler way to find the asymptotic behavior,
%which we explain in A.3.
In A.3, we discuss asymptotic
behavior of the massless scalar correlator in the general
CDL geometry. In A.4, we study the case of a scalar field
which has finite mass in the false vacuum, as a simple
model for realistic moduli fields.

%In this appendix, we calculate the two-point function of the
%minimally coupled massless scalar in 4-D open FRW geometry.
%Following the paper by Gratton and Turok\cite{}, we start from
%Euclidean Coleman De Luccia (CDL) background and analytically
%continue to Lorentzian spacetime. This corresponds to correlation
%function defined on Hawking-Hartle states. In order to get the
%exact results, we study the CDL background in the thin-wall
%approximation. Particularly, we consider the Euclidean De Sitter
%and flat geometry separated by delta-function wall. Then we
%analytically continue the correlation function from Euclidean flat
%region to open FRW region and exam its behavior in the large
%geodesic separation.

\subsection{Euclidean prescription for correlation function}
We calculate the two-point function of a massless minimally
coupled scalar field $\chi$ in the Euclidean CDL geometry
\begin{equation}
 ds^2=a(X)^2(dX^2+d\theta^2+ \sin^2\theta d\Omega_2^2).
\end{equation}
The conformal coordinate $X$ varies over $-\infty<X<\infty$.
The boundary conditions (\ref{boundcond}) tell us that
the scale factor $a(X)$ approaches $c_-e^{X}$ as $X\to -\infty$,
and $c_{+}e^{-X}$ as $X\to \infty$, with some
constants $c_{\pm}$.

It is convenient to consider the rescaled correlator
\begin{equation}
\hat{G}(X_1,X_2;\theta)
%=a(X_1)a(X_2)G(X_1,X_2;\theta)
=a(X_1)a(X_2)\langle
\chi(X_1,0) \chi(X_2,\theta)\rangle,
\end{equation}
which is the correlator of a field $a\chi$ which has
the canonical kinetic term. Using the rotational symmetry we have
brought one of the two points to $\theta=0$, in which case the
correlator depends only on $\theta$ but on no other angular
coordinates.
The correlator $\hat{G}(X_1,X_2;\theta)$ satisfies
\begin{equation}
 \left[-\partial_{X_1}^2+U(X_1)-\nabla^2\right]\hat{G}(X_1,X_2;\theta)
=\delta(X_1-X_2){\delta(\theta)\over \sin^2\theta},
\end{equation}
where $\nabla^2$ is the Laplacian on $S^3$ and $U(X)\equiv
a''(X)/a(X)$.
%where $'=\partial_X$.
We have $U(X)\le 1$ everywhere: It follows from (\ref{xeqs}) that
\begin{equation}
 {a''\over a}-1=-\left({1\over 2}(\Phi')^2+2a^2V(\Phi)\right),
\end{equation}
% we have $a''/a-1=
%-((\Phi')^2/2+2a^2V(\Phi))$,
and the right hand side is negative since
$V(\Phi)\ge 0$ for our case.
The asymptotic value of the potential is
$U(X)\to 1$ as $X\to\pm\infty$.

Let us obtain $\hat{G}(X_1,X_2;\theta)$ in an expansion
in the eigenmodes of $[-\partial_X^2+U(X)]$.
Normalizable modes consist of the continuum modes and the bound
states. The continuum modes $u_k(X)$ satisfy
\begin{equation}
[ -\partial_X^2+U(X)]u_k(X)=(k^2+1)u_k(X)
\end{equation}
with real $k$. These approach the plane wave asymptotically.
The set of orthonormal modes is given by
the waves coming from the left and those coming from the right.
The modes $u_k(X)$ with $k>0$ are defined to be the waves coming
in from the left,
\begin{equation}
 u_k(X)\to e^{ikX}+\rc (k)e^{-ikX}\quad (X\to -\infty), \qquad
u_k(X)\to \tc (k )e^{ikX}\quad (X\to \infty).
\label{asymuk1}
\end{equation}
where $\rc (k)$ and $\tc (k )$ are the reflection and transmission
coefficients; they satisfy $\rc (-k)=\rc ^*(k)$, $\rc (-k)=\tc ^*(k)$,
and $\rc (k)\rc ^*(k)+\tc (k )\tc^*(k)=1$.
The modes $u_{-k}$ $(k>0)$ are the ones coming
in from the right,
\begin{equation}
 u_{-k}(X)\to \tc_r (k)e^{-ikX}\quad (X\to -\infty),\quad
 u_{-k}(X)\to e^{-ikX}+\rc_r (k)e^{ikX}\quad (X\to \infty),
\label{asymuk2}
\end{equation}
where $\rc_r(k)$ and $\tc_r(k)$ are
the reflection and transmission coefficients for the incidence
from the right. In fact, they are related to
%those for the incidence from the left,
$\rc (k)$ and $\tc (k )$ by
$\tc (k )=\tc_r(k)$, $\rc (k)/\rc_r(k)^*=-\tc (k)/\tc (k)^*$.
(See {\it e.g.} \cite{barton} for generalities about
one-dimensional scattering.)

%The modes $u_k(X)$ with $k>0$ are defined by
%\begin{equation}
% u_k(X)\to e^{ikX}+R(k)e^{-ikX}\quad (X\to -\infty), \qquad
%u_k(X)\to \tc (k )e^{ikX}\quad (X\to \infty),
%\label{asymuk1}
%\end{equation}
%and $u_k(X)$ with $k<0$ are defined by
%\begin{equation}
% u_k(X)\to \tc (k )e^{ikX}\quad (X\to -\infty),\qquad
% u_k(X)\to e^{ikX}+R(k)e^{-ikX}\quad (X\to \infty).
%\label{asymuk2}
%\end{equation}
%Here, $R(k)$ and $\tc (k )$ are the reflection and transmission
%coefficients. We have the relations
%$R(-k)=R^*(k)$, $T(-k)=T^*(k)$, and $R(k)R^*(k)+\tc (k )\tc^*(k)=1$.

There is always a zero-energy bound state $u_B(X)\propto a(X)$ which satisfies
\begin{equation}
 [ -\partial_X^2+U(X)]u_B(X)=0.
\end{equation}
In terms of the physical field $\chi$, this mode is simply
\begin{equation}
\chi = constant.
\end{equation}
%This is a mode corresponding to the constant shift of $\chi$.
There is no other bound state, as we will show
in section A.3 by using a technique of supersymmetric quantum mechanics.
In the case of higher dimensional CDL geometry, the corresponding
equation generically has additional bound states.

Using the completeness relation, we rewrite the delta function
\begin{equation}
 \delta(X_1-X_2)=\int_{-\infty}^{\infty}{dk\over 2\pi}
u_k(X_1)u_k^*(X_2)+u_B(X_1)u_B^*(X_2),
%+\sum_{a}\tilde{u}{}_a(X)\tilde{u}{}^*_a(X')
\end{equation}
and we express the Green's function function in the form
\begin{equation}
 \hat{G}(X_1,X_2;\theta)=\int_{-\infty}^{\infty}{dk\over 2\pi}
u_k(X_1)u_k^*(X_2)G_k(\theta)+u_B(X_1)u_B^*(X_2)G_B(\theta)
%+\sum_{a}\tilde{u}{}_a(X)\tilde{u}{}^*_a(X')\tilde{G}_a(\theta)
\label{eucgf}
\end{equation}
where $G_k(\theta)$, $G_B(\theta)$ are Green's
function on $S^3$ with
%masses $k^2+({p\over 2})^2$ and $\lambda_a$, respectively.
appropriate masses;
the Green's function $G_k(\theta)$ satisfies
\begin{equation}
 [-\nabla^2+(k^2+1)]G_k(\theta)
={\delta(\theta)\over \sin^2\theta}~.
\end{equation}
The solution is
\begin{equation}
 G_k(\theta)={\sinh k(\pi-\theta)\over \sin\theta \sinh k\pi}.
\label{gk}
\end{equation}
We can easily confirm (\ref{gk}) by noting that
it is a regular solution of the sourceless equation when
$\theta\neq 0$, and that the singularity at $\theta=0$
is $G_k(\theta)\sim 1/\theta$ independently of $k$.

The Green's function $G_B(\theta)$ corresponding to
the bound state satisfies
\begin{equation}
  -\nabla^2G_B(\theta)={\delta(\theta)\over \sin^2\theta}.
\label{sphgreenfn2}
\end{equation}
We cannot solve this equation, since a constant shift
of $G_B(\theta)$ leaves the right hand side unchanged.
In other words, we cannot have a source in
compact space.

We define the massless Green's function as a limit of the massive
Green's function. The massive Green's function
%(\ref{massivegf})
diverges when we take the massless limit $k\to i$.
We subtract an infinite constant to get a finite Green's
function
\begin{eqnarray}
 G_B(\theta)&=&\lim_{k\to i}\left(G_k(\theta)-{2\over \pi}
{1\over k^2+1}\right)+{1\over 2\pi}\\
&=&{\cos\theta\over \sin\theta}\left(
1-{\theta\over \pi}\right).
\label{masslessgf}
\end{eqnarray}
This prescription for getting the massless Green's function
corresponds to introducing a constant background
charge to make the total charge zero.
%or equivalently to ignoring the zero mode on the sphere.
Since the constant shift
of a massless scalar field does not have physical meaning,
we will have to take derivatives of the correlation function
with respect to both points to get a physical quantity. The above
procedure is equivalent to projecting out the zero mode of the field.

The continuum contribution
to the Green's function (\ref{eucgf}), which we
call $\hat{G}_c(X_1,X_2;\theta)$, can be expressed in a simple form
in the asymptotic limit.
Taking the $X,X'\to -\infty$ limit and using the relations
$\rc (k)^*=\rc (-k)$, $\rc (k)\rc (k)^*+\tc (k )\tc (k )^*=1$, we get
\begin{equation}
 \hat{G}_c(X_1,X_2;\theta)=\int_{-\infty}^{\infty} {dk\over 2\pi}\left(e^{ik\delta{X}}
+\rc (k)e^{-ik\bar{X}}\right)G_k(\theta)
\label{gfcont}
\end{equation}
where $\delta{X}=X_1-X_2$ and $\bar{X}=X_1+X_2$.

We analytically continue the correlator thus obtained
to the Lorentzian space by $X\to T+\pi i/2$, $\theta\to iR$.
The first term in (\ref{gfcont}) gives a
contribution which depends on $\delta T\equiv T_1-T_2$. This piece
exists also in Minkowski space.
The second term in (\ref{gfcont}), which is proportional to $\rc (k)$,
and the bound state contribution depend on $\bar{T}\equiv T_1+T_2$.
These
correspond to the particle production due to a non-trivial geometry.
We will be mostly interested in these latter pieces.

\subsection{Correlation function (a thin-wall example)}

%\subsubsection{Flat space patched to a half sphere}
Let us compute the correlation function explicitly. As an example
of the background, we consider a thin-wall limit where the
Euclidean geometry is a flat space (on the $X\le 0$ side)
patched to a sphere cut in half (on the $X\ge 0$ side).
The scale factor is
\begin{equation}
a(X)=e^X\ \ (X\le 0), \quad a(X)={1\over \cosh X}   \ \ (X\ge 0).
\end{equation}
The geometry in the thin-wall limit is non-analytic at the domain wall,
but the correlator in the FRW region can be obtained by analytic
continuation from the negative $X$ side of the Euclidean geometry.

The potential $U=a''/a$ contains a delta function at the
domain wall $X=0$
\begin{equation}
U(X)= 1- {2\over \cosh^2 X}\Theta(X)-\delta(X),
\end{equation}
and the reflection coefficient is
\begin{equation}
 \rc (k)={i (k+i)\over (2k+i)(k-i)}.
\end{equation}
This $\rc (k)$ can be obtained by matching the wave functions on both sides;
 we will give a simpler derivation in section
A.3 using supersymmetric quantum mechanics.

\subsubsection{Exact expression for the correlator}
Let us calculate the part of the correlator which depends on
$\bar{T}= T_1+T_2$. We first compute the second term of the
continuum contribution (\ref{gfcont}), which depends on $\bar{X}$.
We call this piece $\hat{G}^{(\bar{X})}_c(X_1,X_2;\theta)$. Note
that the asymptotic expression (\ref{gfcont}) is exact in the
$X<0$ region in the thin-wall case. We close the integration
contour in the upper half plane and evaluate the integral as a sum
of the residues from poles. The $1/\sinh k\pi$ factor has poles at
integer multiple of $i$, and $\rc (k)$ has a pole at $k=i$. Thus, in
the upper half plane, we have single poles at $k=2i,3i,\ldots$ and
a double pole at $k=i$.

%As depict in the following figure:

%\begin{figure}[h]
%    \begin{center}
%        \includegraphics[width = 3.5 in,height = 3 in]{Euclidean.eps}
%        \caption{Contour for Euclidean integration. In the upper half
%plane, there are single poles at $k=2i,3i,\ldots$ and
%         a double pole at $k=i$. In the lower half plane, there are
%single poles at $k=-i,
%         -2i,\ldots$ and $k=-i/2$. Note that the position of $k=-i/2$ pole dep%end on the geometry. In general it can be located
%         between $k=0$ and $k=-i$.}
%    \end{center}
%\end{figure}

\begin{figure}[!htb]
\center
%\begin{center}
        \includegraphics[width = 2.5 in,height = 2 in]{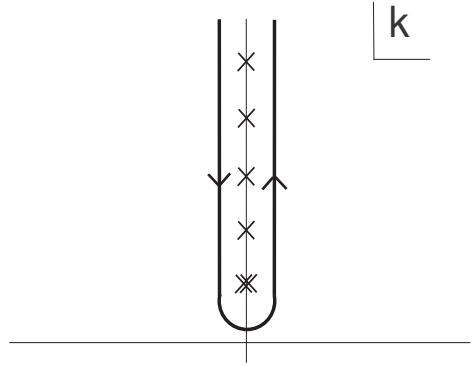}
        \caption{Contour for the integration in (\ref{continuum}). It
surrounds a double pole at $k=i$ and single poles at $k=2i,3i,\ldots$.}
\label{fig-Euclidean}
%Poles in the lower half plane at $k=-i, -2i,\ldots$ and
%at $k=-i/2$ are also depicted; the latter comes from the pole of $\rc (k)$.}
%In the lower half plane, there are
%single poles at $k=-i,
%         -2i,\ldots$ and $k=-i/2$. Note that the position of $k=-i/2$ pole dep%end on the geometry. In general it can be located
%         between $k=0$ and $k=-i$.}
%\end{center}
\end{figure}

%The continuum contribution to the Euclidean Green's function in
%VIS geometry is
The result of the integration along the contour in
Figure~\ref{fig-Euclidean} is
\begin{eqnarray}
&&\hat{G}_c^{(\bar{X})}(X_1,X_2;\theta)
=\int_{-\infty}^{\infty} {dk\over 2\pi}
{i(k+i)\over (2k+i)(k-i)}
e^{-ik\bar{X}}{\sinh k(\pi-\theta)\over
\sin\theta \sinh k\pi}\nonumber\\
&&\qquad=\frac1{2\sin\theta}\frac{2\pi
i}{2\pi}
\sum_{n=2}^{\infty}e^{n\overline{X}}{n+1\over (2n+1)(n-1)}
\frac1{\pi}\left(e^{-in\theta}-e^{in\theta}\right)\nonumber\\
&&\qquad\quad +\frac1{2\sin\theta}\frac{2\pi
i}{2\pi}{(-i)\over \pi}\partial_k\left\{{k+i\over 2k+i}
e^{-ik\bar{X}}(e^{k\pi-k\theta}-e^{-k\pi+k\theta})\right\}\Big|_{k=i}
\label{continuum}\\
&&\qquad ={1\over 12\pi \sin\theta} \Bigg[ -i
e^{(-\bar{X}+i\theta)/2}\log {1+e^{(\bar{X}-i\theta)/2} \over
1-e^{(\bar{X}-i\theta)/2}} +i e^{-(\bar{X}+i\theta)/2}\log
{1+e^{(\bar{X}+i\theta)/2}
\over 1-e^{(\bar{X}+i\theta)/2}}\nonumber\\
&&\qquad \qquad \qquad -4ie^{\bar{X}-i\theta}\log (1-e^{\bar{X}-i\theta})
+4ie^{\bar{X}+i\theta}\log
(1-e^{\bar{X}+i\theta})+\frac23ie^{\bar{X}-i\theta}
-\frac23ie^{\bar{X}+i\theta}\Bigg]\nonumber\\
&&\qquad \quad
+\frac1{2\pi\sin\theta}\left[-\frac{i}9e^{\bar{X}}(e^{-i\theta}-e^{i\theta})-
\frac23(-i\bar{X}+\pi-\theta)e^{\bar{X}-i\theta}+\frac23(-i\bar{X}-\pi+\theta)e^{\bar{X}+i\theta}\right]\nonumber
%&=&{i\over 2\pi\sin\theta}\left\{
%-e^{\bar{X}-i\theta}\log(1-e^{\bar{X}-i\theta})
%+e^{\bar{X}+i\theta}\log(1-e^{\bar{X}+i\theta})\right\}\nonumber\\
%&&-{1\over 2\pi\sin\theta}\left\{
%(-i\bar{X}+\pi-\theta)e^{\bar{X}-i\theta}
%-(-i\bar{X}-\pi+\theta)e^{\bar{X}+i\theta}
%\right\}
\end{eqnarray}

The bound state contribution to the correlator is
%$G_B(X_1,X_2;\theta)$
\begin{equation}
\hat{G}_B(X_1,X_2;\theta)=u_B(X_1)u_B(X_2)G_B(\theta)
={2\over 3}e^{\bar{X}}{\cos\theta\over \sin\theta}\left(
1-{\theta\over \pi}\right),
\label{euclidbs}
\end{equation}
which also depends on $\bar{X}$. The coefficient 2/3
in (\ref{euclidbs}) comes from the normalization of the
bound state wave function $u_B(X)=\sqrt{2\over 3}a(X)$.

We add the continuum contribution (\ref{continuum}) and the bound state contribution (\ref{euclidbs}), and perform the
analytic continuation (\ref{cont}) to the FRW region to get the $\bar{T}$ -dependent part of the rescaled Green's function $\hat{G}^{(\bar{T})}(T_1,T_2;R)$.
%and divide the result
%by factors of $a(X)$ to get the correlator
%$\langle \chi(T,0)\chi(T',R)\rangle$ of a physical field.
The $\bar{T}$ dependent part of the correlator $\langle \chi(T,0) \chi(T',R)\rangle^{(\bar{T})}$
is related to the rescaled Green's function we have calculated
by
\begin{eqnarray}
&&\hspace{-.7cm}\langle \chi(T,0) \chi(T',R)\rangle^{(\bar{T})}=
{\hat{G}^{(\bar{T})}(X_1,X_2;\theta) \over a(T_1) a(T_2)}
%\label{lorentzgf2}\\
\nonumber\\
&&\hspace{-.7cm}=-{2\over 3}i+{R e^R \over 3\pi\sinh R}
+{e^{-\bar{T}}\over 12\pi\sinh R}\Bigg[
ie^{-(\bar{T}+R)/2}\log{1+ie^{(\bar{T}+R)/2}\over
1-ie^{(\bar{T}+R)/2}}
-ie^{-(\bar{T}-R)/2}\log{1+ie^{(\bar{T}-R)/2}\over
1-ie^{(\bar{T}-R)/2}}
\nonumber\\
&&
\hspace{2.5cm}
+4e^{(\bar{T}+R)}\log (1+e^{-(\bar{T}+R)})-4e^{(\bar{T}-R)}\log
(1+e^{(\bar{T}-R)}) +4\bar{T}e^{(\bar{T}-R)}\Bigg]
\label{lorentzgf2}
%+{1\over 2\pi\sinh R}
%\left\{(\bar{T}+R)e^{-R}
%+e^{R}\log(1+e^{-(\bar{T}+R)})
%-e^{-R}\log(1+e^{\bar{T}-R})\right\}.\nonumber
\end{eqnarray}
where the first two terms in the right hand side
do not decay in the large $R$ limit, but rest
of the terms do.

We shall make a technical comment here.
The sum of the double pole contribution and the bound state
contribution, whose Euclidean expressions are given by
the last line of (\ref{continuum}) and by (\ref{euclidbs}),
respectively, is
\begin{equation}
\langle \chi(T_1,0) \chi(T_2,R)\rangle^{(\bar{T})}_{\rm\scriptsize double pole}
+\langle \chi(T_1,0) \chi(T_2,R)\rangle^{(\bar{T})}_B
% \hat{G}^{(\bar{X}),(k=i)}_c(T_1,T_2;R)+\hat{G}_{B}(T_1,T_2;R)
=\frac{1}{\pi}(\frac19-\frac23\pi
i-\frac23\bar{T}),
\end{equation}
and contains only constants and terms linear in $\bar{T}$.
These terms vanish when we take derivatives of
the correlator with respect to both points, so they do
not affect physical quantities. We might call these terms
``pure gauge.''  Up to the pure gauge terms, we can say that
the correlator is given solely by the contributions from
the single poles (at $k=2i,3i,\ldots$).
In fact, that is equivalent to solving for the correlator
as an expansion in the spherical harmonics on $S^3$ (rather
than in the eigenmodes of $[-\partial_X^2+U(X)]$),
leaving out the zero-angular momentum mode on $S^3$.

The $\delta T=T_1-T_2$ dependent part $\langle \chi(T_1,0)
\chi(T_2,R)\rangle^{(\delta T)}$, which might be called the ``flat
space piece,'' is computed from the first term of
(\ref{gfcont}). We get
\begin{equation}
\langle \chi(T_1,0) \chi(T_2,R)\rangle^{(\delta T)}=
-{1\over 2\pi}{e^{-\bar{T}}\over \cosh \delta T -\cosh R},
\label{lorentzgf1}
\end{equation}
for any CDL background. This is simply the usual flat-space correlator for a massless field, written in our coordinate system.

\subsubsection{Large separation behavior}

%study the limit $R\to
%+\infty$, $T\to +\infty$, $(R+T)\to +\infty$.

Let us study the large distance behavior of the correlator.

We have obtained the correlator as a function of $R$, taking one
point to be at the origin of the hyperboloid $R=0$. When the two points
are at general positions, the correlator is a function of
the geodesic distance $\ell$, and is given by replacing
$R$ with $\ell$ in the expressions that we have obtained.
%From now on, we write $\ell$ in place of $R$.
The geodesic distance $\ell$  between the two points on ${\cal H}_3$
located at $(R_1, 0)$ and $(R_2, \alpha)$, where $\alpha$ is an
angle in $S^2$, is given by
%If the two points on $H^3$ are at $(R_1, 0)$ and $(R_2, \alpha)$,
%where $\alpha$ is an angle in $S^2$, the geodesic distance $\ell$
%is given by
\begin{equation}
 \cosh \ell = \cosh R_1\cosh R_2 - \sinh R_1 \sinh R_2 \cos\alpha.
\label{r1}
\end{equation}
This relation can be obtained most easily by considering the inner
product in the embedding space. In the limit of large separation,
(\ref{r1}) reduces to
\begin{equation}
 \ell\sim R_1+R_2+\log (1-\cos\alpha).
\end{equation}
%From now on, we will write $\ell$ in place of $R$.
%With this in mind, in the limit of large geodesic separation,
%(\ref{correlator}) becomes

We take the late time limit as well as the large distance
limit. Namely, we first take the $\bar{T}+\ell\to\infty$ limit
and drop the terms with positive powers of $e^{-(\bar{T}+\ell)}$.
The rest of the terms are given as
an expansion in powers of $e^{\bar{T}-\ell}$.

Let us examine the large distance behavior of the correlator
$\langle \chi \chi\rangle^{(\bar{T})}$, given in
(\ref{lorentzgf2}), in the large distance limit.
We will write $\ell$ in place of $R$.
We will ignore the pure gauge terms.
As we mentioned, there is a piece in $\langle \chi \chi\rangle^{(\bar{T})}$
that does not decay at large distance. We shall call it $G_2$,
\begin{equation}
 G_2= {\ell e^\ell\over 3\pi\sinh \ell}.
\end{equation}
This grows linearly with $\ell$, and in terms of the
coordinates, its asymptotic form is
\begin{equation}
 G_2 \sim R_1+ R_2 +\log (1-\cos\alpha).
\end{equation}
The the first two terms, $R_1$ and $R_2$, do not affect
physical quantities because they disappear when we take derivatives with respect to the locations of both points. The last term, however, is physical. This piece
represents a correlation which is independent of
the radial position. Its interpretation is discussed further in the main text.

To find the subleading terms, we note that the
flat space piece $\langle \chi \chi\rangle^{(\delta T)}$
in (\ref{lorentzgf1}) yields
%a $\delta T$ independent term
$e^{-\bar{T}-R}/\pi$ when expanded
in the large $R$ limit. This term is independent
of $\delta T$, and precisely cancels the order $e^{-R}$
piece in $\langle \chi \chi\rangle^{(\bar{T})}$.
We call $G_1$ the sum of the piece of $\langle \chi \chi\rangle^{(\bar{T})}$
apart from $G_2$ and $e^{-\bar{T}-R}/\pi$ from $\langle \chi \chi\rangle^{(\delta T)}$,
\begin{equation}
 G_1=\frac{1}{12\pi}e^{-(\bar{T}+\ell)}\left\{-\frac23e^{\bar{T}-\ell}
+O(e^{2(\bar{T}-\ell)})
 -\pi e^{-(\bar{T}+\ell)/2}+O(e^{-2(\bar{T}+\ell)})\right\}.
%+\frac2{3\pi}e^{\bar{T}}(\ell-i\pi)
\label{g1asympt}
\end{equation}
This is an expansion in powers of $e^{\bar{T}-\ell}$ and of
$e^{-(\bar{T}+\ell)}$. Higher orders of the expansions are
indicated by $O(e^{2(\bar{T}-\ell)})$ and $O(e^{-2(\bar{T}+\ell)})$.
In the limit of our interest, we ignore the last two terms of
(\ref{g1asympt}) which have positive powers of $e^{-(\bar{T}+R)}$.
The leading term, which is the first term of (\ref{g1asympt}),
goes as $e^{-2\ell}$. We will call it the ``dimension 2 piece.''

%(From now on, we write $\ell$ in place of $R$.)
%As we mentioned,
%\begin{eqnarray}
%\hat{G}(T_1,T_2;\ell)&=&\frac{e^{\bar{T}}}{12\pi}
%e^{-(\bar{T}+\ell)}\Big(-\frac23e^{\bar{T}-\ell}
%+4(\bar{T}+\ell)e^{\bar{T}-\ell}+O(e^{2(\bar{T}-\ell)})\nonumber\\
%&&-\pi e^{-(\bar{T}+\ell)/2}+O(e^{-2(\bar{T}+\ell)})\Big)
%+\frac2{3\pi}e^{\bar{T}}(\ell-i\pi)
%\end{eqnarray}

%Now we consider the behavior of the correlator near the spatial
%infinity in the late time limit. Namely, we take the
%$\bar{T}+R\to\infty$ limit first and ignore terms which have
%positive powers $e^{-(\bar{T}+R)}$, and then take the
%$R-\bar{T}\to\infty$ limit and get the correlator
%in an expansion in powers of $e^{\bar{T}-R}$.

\subsubsection{Asymptotic behavior and the poles of $\rc (k)$}
Each term in the series in (\ref{g1asympt}) can be interpreted
as arising from the poles of $\rc (k)G_k(\theta)$. We shall give an alternative
derivation of the asymptotic behavior (\ref{g1asympt}).

%We can understand this asymptotic behavior in an alternative
%way. Each term in (\ref{g1asympt}) can be directly interpreted
%as a contribution from the poles of $\rc (k)$.

The Euclidean expression for the correlation function
($\bar{X}$ dependent part) is
\begin{equation}
\langle \chi(X_1,0)\chi(X_2,\theta)\rangle^{(\bar{X})}=
e^{-\bar{X}}\oint_C {dk\over 2\pi}
\rc (k)e^{-ik\bar{X}} {\sinh k(\pi-\theta)\over \sin \theta \sinh k\pi}
\end{equation}
where the contour $C$ surrounds the single poles at $k=2i,3i,\ldots$.
This contour is depicted in the main text, section \ref{sec-corr}, figure~\ref{fig-Lorentz}~(a).
As we have mentioned, we can define the correlator either as
the sum over contributions from poles at $k=i,2i,\ldots$ plus
the bound state contribution, or as the contributions from the
single poles alone. We take the latter definition here\footnote{If
we include the $k=i$ pole
and the bound state in the definition of the Euclidean
correlator and do the argument below, we will conclude
that the linearly growing piece comes from the bound
state, and dimension 2 piece comes form the bound state
and the $k=i$ pole.}.

We perform analytic continuation at this stage,
\begin{equation}
\langle \chi(T_1,0)\chi(T_2,R)\rangle^{(\bar{T})}
=e^{-\bar{T}}\oint_C {dk\over 2\pi}
\rc (k)e^{-ik\bar{T}}{ \left(e^{-ikR}-e^{ikR-2\pi k}\right)
\over 2\sinh R \sinh k\pi}.
\label{contourint}
\end{equation}
If $+\bar{T}<0$ and $R-\bar{T}>0$, the integral along
the contour $C$ converges.
In the region of interest $R+\bar{T}>0$, $R-\bar{T}>0$,
the second term containing $e^{ik(R-\bar{T})}$
converges along $C$, but the first term containing
$e^{-ik(R+\bar{T})}$ does not. The correct
analytic continuation
of the $e^{-ik(R+\bar{T})}$ term is given by the integration
along a contour closed in the opposite direction,
which is depicted in Figure~\ref{fig-Lorentz}~(b).
%as a sum of the residues of the poles,
%we have to close the
%integration contour in the upper half plane for the
%terms containing $e^{ik(\bar{T}-R)}$, and in the
%lower half plane for those
%containing $e^{-ik(\bar{T}-R)}$.
The summation over the contributions from the poles takes
the form of an expansion valid in the large $R$ limit.

%\begin{figure}[h]
%    \begin{center}
%        \includegraphics[angle = 90,width = 5.5 in,height = 3 in]{Lorentz.eps}
%        \caption{(a) is the contour for the first term and (b) is the contour %for the second term. Note that the pole
%        at $k=0$ only appears if we separate two terms.}
%    \end{center}
%\end{figure}

The integration for the $e^{ik(R-\bar{T})}$ terms
picks up the residues of the single poles
at $k=2i,3i,\ldots$, and gives rise to terms of the form
$e^{-\bar{T}}e^{-n(\bar{T}-R)}/\sinh R$ with $n=2,3,\ldots$.
These terms are subleading compared to the most interesting terms; as explained in the main text, each term takes the form of the correlator in Euclidean AdS$_3$ with a given $m^2$, and thus has a definite dimension in the CFT.

The integration for the $e^{-ik(\bar{T}+R)}$ terms
along the contour
in Figure~\ref{fig-Lorentz}~(b) first picks up the contribution
from the double pole at $k=i$. It is of the form
\begin{equation}
 {ie^{-\bar{T}}\over 2\pi \sinh R}\partial_k
\left\{(k-i)\rc (k)e^{-ik(\bar{T}+R)}\right\}\Big|_{k=i}.
\end{equation}
This pole yields the terms of interest,
the term that linearly grows with $R$ and the
$e^{-2R}$ term, plus pure gauge terms.

There are single poles at $k=0,-2i,-3i,\ldots$, and also
at $k=-i/2$ from the pole of $\rc (k)$. There is no pole
at $k=-i$, because $\rc (k)$ generically has a zero at $k=-i$,
as we show in section A.3.

From the pole at $k=0$, we get
\begin{equation}
 {-ie^{-\bar{T}}\over 2\pi\sinh R}\rc (k=0).
\label{kzeropole}
\end{equation}
Here $\rc (k=0)=-1$ generally, except for the
perfectly reflectionless potential for which $\rc (k)=0$ identically.
(See {\it e.g.} \cite{barton}.)
The piece (\ref{kzeropole}) is
canceled by the $k=0$ pole contribution from the
flat space piece,
\begin{equation}
e^{-\bar{T}}\int {dk\over 2\pi}
e^{ik\delta T}{ \left(e^{-ikR+k\pi}-e^{ikR-k\pi}\right)
\over 2\sinh R \sinh k\pi}.
\end{equation}
Note that though the sum of the two terms in the integrand
is regular at $k=0$, each term separately has a pole.
We pick up this pole, since we are closing the contour
in the different direction for each term.
%which is independent of $\delta T$.

The rest of the contributions from the integral of
the $e^{-ik(\bar{T}+R)}$ term are of the form
$e^{-\bar{T}}e^{n(\bar{T}+R)}/\sinh R$
with negative $n$. These terms can be neglected in our
limit.

%We have seen that the poles of $\rc (k)$ in the upper half
%plane gives the contributions relevant in our limit,
%and that the $k=i$ pole gives the leading pieces.

\subsection{Asymptotic behavior in the general CDL geometry}

We have studied the large distance limit of the correlator
for a particular background, and found a piece
which linearly grows with $\ell$ and a `dimension 2' piece
which goes like $e^{-2\ell}$. We now show that this behavior
is true for general CDL backgrounds.
As we have mentioned, the asymptotic behavior is determined
by the poles of the reflection coefficient $\rc (k)$. In this
subsection, we obtain $\rc (k)$ for general thin-wall geometries,
and show that the above asymptotic behavior is general.
Then we argue that our statement about the asymptotic behavior
is true for thick-wall cases as well.

%is true in general thin-wall. We then argue that the pole
%structures of non thin-wall cases are similar to the thin-wall
%limit.

%Since the asymptotic behavior
%is determined from the reflection coefficient, we shall
%first obtain the reflection coefficient
%
%The asymptotic behavior found in the last subsection
%for an example is true in the general CDL geometry.
%
%As we have shown, the large distance behavior of
%our interest is determined by the poles of the reflection
%coefficient in the upper half plane. We will first obtained
%the reflection coefficient for the general thin-wall cases.

\subsubsection{Reflection coefficient for general thin-wall}
We consider general thin-wall geometries, where  the size of
the critical bubble (size of the bubble at the moment of nucleation)
and the de Sitter radius are arbitrary. The Euclidean geometry is
a flat disc patched to an arbitrary portion of a sphere. The
scale factor is
\begin{equation}
 a(X)={e^{X-X_0}\over \cosh X_0}\ \ (X\le X_0),\qquad
 a(X)={1\over \cosh X}\ \ (X\ge X_0),
\label{generala}
\end{equation}
where we have set the de Sitter radius to unity.
The $X_0\to -\infty$ limit corresponds to the small
critical bubble, so that the Euclidean geometry includes almost all of the de Sitter sphere.
 The $X_0\to +\infty$ limit
corresponds to the case where the Euclidean geometry includes only a tiny piece of the de Sitter sphere.
%Geometry in the latter limit is equivalent to the VIS geometry
%under overall rescaling of the geometry.
The Green's function can be calculated as in section A.1, A.2.
The potential $U(X)=a''/a$, is given by
\begin{equation}
 U(X)=1-{2\over \cosh^2X}\Theta(X-X_0)-{e^{X_0}\over \cosh X_0}
\delta (X-X_0).
\end{equation}

To find the reflection coefficient, it is useful to note
that the potential of the form $U(X)=a''/a$ can be
embedded in supersymmetric quantum mechanics. (See
\cite{susyqm} for a review on supersymmetric
quantum mechanics.)
The potential can be written as
\begin{equation}
 U(X)=(W')^2+W''
\end{equation}
with $W=\log a$. We define the partner potential $\tilde{U}(X)$ by
\begin{equation}
 \tilde{U}(X)=(W')^2-W''.
\end{equation}
The Schr\"{o}dinger operator $(-\partial_X^2+U)$ for the original potential
is written as $(\partial_X+W')(-\partial_X+W')$, and that
for the partner potential is written as
$(-\partial_X+W')(\partial_X+W')$.
The wave function $u_k$ for $U(X)$ is related to the one $\tilde{u}_k$
for $\tilde{U}(X)$ with the same energy eigenvalue by the supersymmetry
transformation $\tilde{u}_k=(-\partial_X+W')u_k$.
%We can see it from
There is a one to one correspondence of the states except for the zero
energy ground state $u_B=a(X)$ for $U(X)$, which
is annihilated by the transformation $(-\partial_X+W')u_B=0$.

The partner potential is flat either for the
flat space potential $U(X)=1$, or for the sphere (de Sitter) potential
$U(X)=1-2/\cosh^2 X$. Thus,  $\tilde{U}(X)$ for the thin-wall case
is just a repulsive delta function at $X=X_0$:
\begin{equation}
 \tilde{U}(X)=1+{e^{X_0}\over \cosh X_0}\delta(X-X_0).
\end{equation}
The reflection coefficient $\tilde{\rc }(k)$ for $\tilde{U}(X)$ is
obtained by matching the plane waves on the two sides,
\begin{equation}
 \tilde{\rc }(k)
%={-i {e^{X_0}\over \cosh X_0}e^{2ikX_0}
%\over 2k+i {e^{X_0}\over \cosh X_0}}
={-ie^{2ikX_0}\over (e^{-2X_0}+1)k+i}.
\label{eq:tilder}
\end{equation}
Since the potential is repulsive, there is no bound state, and
$\rc (k)$ does not have a pole in the upper half $k$ plane.
%A pole of $\rc (k)$ in the upper half $k$ plane corresponds to
%a (normalizable) bound state.

Reflection and transmission coefficients for the original potential,
$\rc (k)$ and $\tc (k )$, are defined in (\ref{asymuk1}).
Those for the partner potential $\tilde{\rc }(k)$ and $\tilde{T}(k)$
are defined similarly from the asymptotic form of $\tilde{u}_k$.
%\[
% \psi_k \sim e^{ikX}+\rc (k)e^{-ikX}\ \ (X\to -\infty),\quad
%\psi_k \sim \tc (k )e^{ikX} \ \ (X\to \infty),
%\]
%and similarly for $\tilde{\rc }(k)$, $\tilde{T}(k)$, we have
%the following relations
Applying the supersymmetry transformation to the asymptotic form
of $u_k$, we get the relation
\begin{equation}
\tilde{\rc }(k)={ik+W'_-\over -ik+W'_-}\rc (k),\quad
\tilde{T}(k)={-ik+W'_+\over -ik+W'_-}\tc (k ),
\label{eq:tildert}
\end{equation}
where $W'_\pm=W'(\pm\infty)$. Since the scale factor
$a(X)$ behaves as  $a(X)\propto e^{\pm X}$
for $X\to \mp \infty$, we always have $W'_\pm=\mp 1$, and
\begin{equation}
%\tilde{\rc }(k)=-{k-i\over k+i}\rc (k),\quad
%\tilde{T}(k)={k-i\over k+i}\tc (k ).
\rc (k)=-{k+i\over k-i}\tilde{\rc }(k),\quad
\tc (k )={k+i\over k-i}\tilde{T}(k).
\label{tildert2}
\end{equation}
From (\ref{eq:tilder}), we get
\begin{equation}
 \rc (k)={i(k+i)e^{2ikX_0}\over (k-i)((e^{-2X_0}+1)k+i)}.
\label{eq:rkgeneral}
\end{equation}

As we mentioned in section A.2, the bound state pole
at $k=i$ is responsible for the leading terms in
the large distance limit that we are interested in.
We found here that there is no other poles than at $k=i$
in the upper half plane. In the lower half plane,
there is a pole at $k=-(e^{-2X_0}+1)^{-1}i\equiv ai$
and a zero at $k=-i$. The pole is
located between $k=0$ and $k=-i$ on the imaginary axis
(in the half sphere example, at $k=-i/2$). This pole
gives rise to the term $e^{-a({\bar{T}+R})}$, which
can be neglected since we take the $\bar{T}+R\to \infty$
limit first.

The above discussion can be also applied to thick-wall cases.
The potential is of the supersymmetric form. The
partner potential is generally repulsive, since
\begin{equation}
 \tilde{U}(X)=-{a''\over a}+2\left({a'\over a}\right)^2
={3\over 2}\left(\Phi'\right)^2+1\ge 1
\end{equation}
as we see from the Euclidean FRW equations (\ref{xeqs}).
Thus, the partner potential will not have a bound state,
and the reflection coefficient $\tilde{\rc }(k)$ will not have
a pole in the upper half plane\footnote{There could be
resonance poles,
%away from the imaginary axis,
but they must be in the lower half plane and cannot affect
the behavior of the correlator in our limit.
Eigenvalue $k^2+1$ of the Schr\"{o}dinger operator
$[-\partial_X^2+U(X)]$ cannot be
complex unless the self-adjointness of the operator is
broken. In other words, the wave function for the
resonance should blow up at $X\to \pm\infty$, thus
$k$ is in the lower half plane.}.
The reflection coefficient for the original potential $\rc (k)$
is still related to $\tilde{\rc }(k)$ by (\ref{tildert2}),
since this relation was derived using only the asymptotic form
of $a(X)$.  This suggests that $\rc (k)$ has only one pole
in the upper half plane at $k=i$, and that the asymptotic
behavior of the correlator that we found in section A.2
is true in general.

\subsection{Scalar with mass in the false vacuum}

So far, we have been considering a massless scalar, but
we do not expect it to be present in the realistic
situations. The ``moduli field'' $\Phi$ might be massless
in the true vacuum, but it should be stabilized in the
false vacuum.

%We have been studying the massless scalar in the CDL geometry
%as a model for the graviton. Actual scalar may be massless in
%the FRW region, but they should have
%mass in the false vacuum in general.

To understand the behavior of scalars which have mass in the
false vacuum, we shall study a simplified model.
%To understand the qualitative behavior of actual scalar field,
%we consider a limit where the mass is zero
%in the flat region and infinite in the de Sitter region.
We consider a scalar which is massless in the flat space, and
has infinite mass in the de Sitter space.
Infinite mass makes the potential infinitely high.
Wave function cannot penetrate into the de Sitter
region, and the reflection coefficient is $\rc (k)=-e^{2ikX_0}$,
where the real number
$X_0$ is the position of the domain wall as before.
There is no bound state.

The correlation function is calculated from (\ref{gfcont}) and
is given by
\begin{equation}
 \langle \chi(T_1,0) \chi(T_2,R) \rangle =
-e^{2X_0}\cosh^2 X_0
{e^{-\bar{T}}\over 2\pi}\left(
{1\over \cosh(\bar{T}-2X_0)+\cosh R}+{1\over \cosh \delta T
-\cosh R}\right),
\end{equation}
where the factors outside the bracket are due to the rescaling by
$a^{-1}(T_1)a^{-1}(T_2)$ to get the correlator of the physical field
$\chi$. This correlator does not contain a piece which grows in the
large distance. The leading term has dimension 2,
%The leading term in the limit of $R\to \infty$, $\bar{T}\to \infty$,
%$\delta{T}<<\bar{T}$ is the dimension 2 piece,
\begin{equation}
 \langle \chi(T_1,0) \chi(T_2,R) \rangle\sim
 \cosh^2 X_0{e^{-2R}\over \pi}.
\end{equation}
%The magnitude of this term becomes large in the small bubble
%limit $X_0\to -\infty$.  This is also large in the $X_0\to +\infty$
%limit, this does not mean that there is a large physical effect,
%since it dissapears if we rescale the length scale to make
%the size of the Euclidean geometry order one.

When we have small but finite mass, there will be no linearly
growing term, but still there will be non-normalizable correlations
which scale like $e^{-\epsilon R}$. We expect this to happen as long
as the bound state exists. Note that the effect of mass is to make
the potential $U(X)$ shallow by adding a term $a(X)^2m^2$. Up to
certain value of mass, there will be a bound state. The bound state
energy should be in the range $0\le (k^2+1)\le 1$ (smaller than the
asymptotic value of the potential); the bound state pole will be
shifted from $k=i$ to $k=(1-\epsilon)i$, with $0\le \epsilon \le 1$.

The Euclidean correlator is expressed as an integral similar to the
one for the massless case. Now the correlator is defined by an
integral along the contour surrounding the poles at $k=i,2i,\ldots$,
which goes between the poles at $k=i$ and $k=(1-\epsilon)i$. As in
the massless case, the analytic continuation to Lorentzian is done
by closing the contour for the term proportional to
$e^{-ik(\bar{T}-R)}/\sinh R$ in the upper half plane, and the
contour for the term proportional to $e^{-ik(\bar{T}+R)}/\sinh R$ in
the lower half plane. The former integration picks up a contribution
from the $k=i$ pole, which is now a single pole. It gives a
dimension 2 piece. The latter integration has a contribution from
the $k=(1-\epsilon)i$ pole. It gives a non-normalizable dimension
$\epsilon$ piece, along with a subleading dimension $2+\epsilon$
piece.

It is instructive to consider the massless ($\epsilon\to 0$) limit.
The sum of the $k=i$ and the $k=(1-\epsilon)i$ pole contributions is
of the form \be {1\over \epsilon} \left({e^{-R}\over \sinh R}
-{e^{(1-\epsilon)R}\over \sinh R}\right). \ee The $1/\epsilon$
factor reflects the fact that the residue at these poles are large
because of the presence of the nearby pole. This factor has a
relative minus sign for the two poles. In the $\epsilon\to 0$ limit,
we recover the massless correlator, up to a divergent constant: The
linearly growing term appears by taking the limit of $e^{\epsilon
R}/\epsilon$. We also see that the $Re^{-2R}$ term in the massless
correlator is a result of the coalescence of the dimension 2 piece
and the dimension $(2+\epsilon)$ piece which have opposite signs.

%Up to certain value of mass, there will be a bound state. The bound
%state energy $\epsilon^2$ could be $0\le\epsilon^2<1$ (smaller than
%the asymptotic value of the potential).
%In this case, the correlator
%calculated similarly to the the previous sections will have a term
%which goes like $e^{-\epsilon R}$. When the mass is so large that
%there is no bound state, the leading term will be the $e^{-2R}$
%term.

%When the mass is smaller than a threshold, there will still be a
%non-normalizable piece, which scales like $e^{-\epsilon R}$ with
%$0<\epsilon <1$ in the large $R$ limit. Although it does not grow,
%it does not decay fast enough to be normalizable.

%%%%%%%%%%%%%%%%%%%%%%%%%%%%%%%%%%%%%%%%%%%%%%%%%%%%%%%%%%%%%%%
%%%%%%%%%%%%%%%%%%%%%%%%%%%%%%%%%%%%%%%%%%%%%%%%%%%%%%%%%%%%%%%
%%%%%%%%%%%%%%%%%%%%%%%%%%%%%%%%%%%%%%%%%%%%%%%%%%%%%%%%%%%%%%%

\section{Graviton correlator in the CDL geometry}
In this appendix we calculate graviton correlators in
the CDL geometry. As in the scalar case, we get
the correlator in the Euclidean CDL geometry
and then analytically continue it to Lorentzian signature.

Our method explained in section B.1 and B.2 closely follows
the one of Hawking, Hertog and Turok. See
\cite{turok1, turok2} for more details. (See also
\cite{openinflation, sasaki}.) However, in section B.3,
we reach a conclusion on the non-normalizable mode different
from theirs.

\subsection{General prescription}
We consider the transverse traceless (TT) mode of graviton
in the Euclidean CDL geometry.
We define $h_{ab}$ to be the
fluctuation of the metric around the CDL background
$g_{ab}\to g_{ab}^{(CDL)}+h_{ab}$ which is TT on the
$S^3$: $h^{a}_{a}=\nabla_{a}h^{a}_{b}=0$. We will
use $a,b$ to denote the indices on $S^3$, or on $\hyp$
after analytic continuation.

%We consider a mode of graviton in the Euclidean CDL geometry
%which is transverse and traceless (TT) on $S^3$:
%$h^{a}_{a}=\nabla_{a}h^{a}_{b}=0$. We use $a,b$ to denote
%the indices on $S^3$.
The linearized equation of motion for the TT graviton
(rescaled by $a(X)$) is
\begin{equation}
[ -\partial^2_{X} +U(X)+2-\nabla^2](a(X)h_{ab}(X,\theta))=0,
\label{heom}
\end{equation}
where $U(X)=a''(X)/a(X)$.
The eigenvalue of the Laplacian on $S^3$ for the TT tensor
is
$\nabla^2h_{ab}=[-l(l+2)+2]h_{ab}$ with $l=2,3,\ldots$.
The equation of motion (\ref{heom}) is the same as
%the one for the massless minimally coupled scalar field,
the massless minimally coupled scalar equation,
except that the angular momentum $l$ starts from 2 rather than 0.
%except
%for the range of $l$; in the scalar case, $l$ starts from
%$l=0$.
%(We use the same symbol $l$ for
%geodesic distance; it should be clear from the context
%which one is meant.)

We want to find the correlator
%We want to find the correlator of the TT mode of the graviton
$\hat{G}^{ab}{}_{a'b'}(X_1,X_2;\theta)=a(X_1)a(X_2)\langle h^{ab}(X_1,\theta)
h_{a'b'}(X_2,0)\rangle$, which satisfies
\begin{equation}
[ -\partial^2_{X_1} +U(X_1)+2 -\nabla^2]\hat{G}^{ab}{}_{a'b'}
(X_1,X_2;\theta)=\delta(X_1-X_2){\delta(\theta)\over \sin^2\theta}
\delta^{ab}{}_{a'b'}
\end{equation}
where $\delta^{ab}{}_{a'b'}=\delta^{a}_{a'}\delta^{b}_{b'}
+\delta^{a}_{b'}\delta^{b}_{a'}-{2\over 3}
g^{ab}g_{a'b'}$.

The correlator can be expanded into harmonics on $S^3$,
\begin{equation}
 \hat{G}^{ab}{}_{a'b'}(X_1,X_2;\theta)=
\sum_{k=3i}^{+i\infty} \tilde{G}_{k}(X_1,X_2)W^{ab}_{(k)a'b'}(\theta)
\label{gexpand}
\end{equation}
where the sum is over $k=i(l+1)$ with $l=2,3,\ldots$. The bitensor
(tensor with indices defined at two separate points)
$W^{ab}_{(k)a'b'}(\theta)$ is an eigenfunction of the Laplacian
\begin{equation}
 \nabla^2 W^{ab}_{(k)a'b'}(\theta)=(k^2+3)W^{ab}_{(k)a'b'}(\theta),
\end{equation}
and is regular over $S^3$ when $k$ takes the above values.
It is a sum of products of tensor harmonics
$q^{(k)}_{{\cal P}nm}{}^{ab}$ at two points ($\theta$ and 0);
we fix $k$ and sum over other
the quantum numbers ${\cal P}$, $m$, $n$ (parity and $S^2$ angular
momenta),
\begin{equation}
 W^{ab}_{(k)a'b'}(\theta)=\sum_{{\cal P}nm} q^{(k)}_{{\cal P}nm}{}^{ab}(\theta)
q^{(k)}_{{\cal P}nm,}{}_{ab}(0)^*,
\end{equation}
where the tensor harmonics satisfy the eigenvalue equation
%$q^{(k)}_{{\cal P}nm}{}^{ab}$
$\nabla^2 q^{(k)}_{{\cal P}nm}{}^{ab}=(k^2+3)q^{(k)}_{{\cal P}nm}
{}^{ab}$, and the normalization
$\int d\Omega q^{(k)}_{{\cal P}nm}{}^{ab} q^{(k')*}_{{\cal
P}'n'm',}{}_{ab}
=\delta^{kk'}\delta_{{\cal P}{\cal P}'}\delta_{nn'}\delta_{mm'}$.
There is a completeness relation
\begin{equation}
 \sum_{k=3i}^{+i\infty}W^{ab}_{(k)a'b'}(\theta)
={\delta(\theta)\over\sin^{2}\theta}\delta^{ab}{}_{a'b'}.
\end{equation}
%and ${\cal P}$, $m$, $n$ denotes the parity,
%and angular momenta on $S^2$.
The bitensor $W^{ab}_{(k)a'b'}(\theta)$ is invariant under the $SO(4)$
isometry, and is called
the ``maximally symmetric bitensor''~\cite{allen, allenjacobson}.
As in the scalar case, we will
regard $k$ as a general complex number and
obtain the correlator in an integral representation.

%is called the maximally symmetric bitensor on $S^3$.
%is regular over $S^3$ when
%$k=i(l+1)$ with $l=2,3,\ldots$,
%$W^{ij}_{(k)i'j'}(\theta)$
%
%and the sum in (\ref{gexpand}) is over these values of $k$.
%It is useful to regard $k$ as a complex number since
%the large distance behavior of the correlator is determined
%by the pole structure of
%$G_{k}(X,X')$ and $W^{ab}_{(k)a'b'}(\theta)$ as functions of $k$.
%
%The bitensor $W^{ab}_{(k)a'b'}(\theta)$
%is an eigenfunction of the Laplacian
%\begin{equation}
% \nabla^2 W^{ab}_{(k)a'b'}(\theta)=(k^2+3)W^{ab}_{(k)a'b'}(\theta).
%\end{equation}
%The bitensor $W^{ij}_{(k)i'j'}(\theta)$
%It is a sum of products of tensor harmonics at the two points
%($\theta$ and $0$) with the given eigenvalue, and
%it is invariant under the isometry $SO(4)$.

The function $\tilde{G}_{k}(X_1,X_2)$ satisfies
\begin{equation}
 \left[-\partial_{X_1}^2+U(X_1)-(k^2+1)\right]\tilde{G}_k(X_1,X_2)=\delta(X_1-X_2).
\label{gxx}
\end{equation}
%The Green's function $G_{k}(X,X')$ can
The solution can be written in terms of
solutions of the homogeneous equation,
\begin{equation}
\tilde{G}_k(X_1,X_2)
=\frac1{\Delta_k}\left[u_k^R(X_1)u_k^L(X_2)\Theta(X_1-X_2)
+u_k^L(X_1)u_k^R(X_2)\Theta(X_2-X_1)\right]
\label{tildegxx}
\end{equation}
where $u_k^R(X)$ and $u_k^L(X)$ are
the solution of (\ref{gxx}) without
the delta function on the right hand side;
$u_k^R(X)$ approaches $e^{ikX}$ as $X \to +\infty$, and
$u_k^L(X)$ approaches $e^{-ikX}$ as $X \to -\infty$.
$\Delta_k=-2ik$ is the Wronskian of $u_k^L$ and $u_k^R$.
Since ${\rm Im} k>0$ here, (\ref{tildegxx}) is regular away from the
coincident point.
When $X_1,X_2\to -\infty$ and $X_2< X_1$, (\ref{tildegxx})
becomes\footnote{
The procedure described here completely parallels the one
explained in section A.1.
If we apply this method to a scalar,
we will have a sum of the product of scalar harmonics
$\sum q^{(k)}(\theta) q^{(k)}(0)^*
\propto k \sinh k\theta/ \sin \theta $, in place of the
bitensor, and  we will have the same integrals as
the ones in section A.1.}
\begin{equation}
 \tilde{G}_{k}(X_1,X_2)={i\over 2k}\left(e^{ik\delta X}+\rc (k)e^{-ik\bar{X}}
\right),
\label{gxasympt}
\end{equation}
where $\delta X=X_1-X_2$ and $\bar{X}=X_1+X_2$.

The graviton correlator in the Lorentzian CDL geometry is
given by analytically continuing (\ref{gexpand}).
We rewrite the sum (\ref{gexpand}) as an integral over $k$
before analytic continuation, and study the large distance
behavior of the Lorentzian correlator using the contour
deformation argument explained in A.2.3. The sum (\ref{gexpand})
can be written as
\begin{equation}
 \hat{G}^{ab}{}_{a'b'}(X_1,X_2;\theta)=\oint_{\cal C} {dk\over 2 i}
\coth k\pi \,\tilde{G}_{k}(X_1,X_2)W^{ab}_{(k)a'b'}(\theta).
\end{equation}
The factor $\coth k\pi$ gives rise to poles at integer
multiple of $i$. The contour ${\cal C}$ is the one
shown in Figure~\ref{fig-Lorentz}~(a) which surrounds
$k=2i, 3i, 4i, \ldots$. In fact, for the tensor case,
we do not have a pole at $k=2i$
%and gives the contributions at
%these values of $k$.
%The sum (\ref{gexpand}) starts from $k=3i$,
since the bitensor is zero there,
$W^{ab}_{(k=2i)a'b'}(\theta)=0$~\cite{allen, turok2}.

%We continue
%$X\to T+{\pi\over 2}i$ and $\theta\to i\ell$, where $\ell$
%is the geodesic distance on the hyperboloid.
%Rather than performing
%the summation explicitly, it is more convenient to rewrite
%(\ref{gexpand}) as an integral over $k$
%\begin{equation}
% \hat{G}^{ab}{}_{a'b'}(X_1,X_2;\theta)=\int_{\cal C} {dk\over 2 i}
%\coth k\pi \,\tilde{G}_{k}(X_1,X_2)W^{ab}_{(k)a'b'}(\theta).
%\end{equation}
%
%The factor $\coth k\pi$ gives rise to poles at integer
%multiples of $i$. The contour ${\cal C}$ surrounds the poles
%at $k=3i, 4i, \ldots$.

We perform analytic continuation to Lorentzian signature by
$X\to T+{\pi\over 2}i$ and $\theta\to i\ell$, where $\ell$
is the geodesic distance on the hyperboloid.
\begin{equation}
  \hat{G}^{ab}{}_{a'b'}(T_1,T_2;\ell)
=\int {dk\over
k\sinh k\pi} \, \left(e^{ik\delta T}+\rc (k)e^{-ik\bar{T}}
\right) W^{ab}_{(k)a'b'}(\ell).
\label{gravlorentz}
\end{equation}
As in the scalar case in A.2.3, for some terms,
this integral along the coutour
${\cal C}$ does not converge in the region of interest
$\ell+\bar{T}\to \infty$, $\ell-\bar{T}\to\infty$.
The correct analytic continuation for these terms is given by
closing the coutour in the other direction.
Let us first find the explicit form the
Lorentzian bitensor $W^{ab}_{(k)a'b'}(\ell)$ in the
next subsection.
%In the next subsection,
%We will study its large distance behavior
%in the same way as we did in the scalar case in section A.2.3.
%As we see in the next subsection, bitensor consists of terms
%involving $e^{ik\ell}$ and $e^{-ik\ell}$.
%To evaluate the integral as a sum of residues,
%we have to close the contour in the
%upper half plane for the former terms,
%and  in the lower half plane for the latter terms.
%and in the lower half plane for those involving
%for others. As we have explained in section A.2.3,

\subsection{The Lorentzian bitensor}
We first present the general expression in terms
of the geodesic distance obtained in~\cite{allen}. Then
we evaluate it on the geodesic which connects two
points near the boundary.
\subsubsection{General form}
The maximally symmetric bitensor $W^{ab}_{(k)a'b'}(\ell)$
%on $H^3$, which is given by the analytic continuation of the one on $S^3$,
%takes the form
has the following tensor structure,
%\begin{eqnarray}
%W^{(k)}_{iji'j'}(\ell)\!
%&\!=&\!w_1^{(k)}(\ell)[g_{ij}-3n_in_j][g_{i'j'}-3n_{i'}n_{j'}]\nonumber\\
%&&\!\! +w_2^{(k)}(\ell)[C_{jj'}n_in_{i'}+C_{ij'}n_jn_{i'}
%+C_{ji'}n_in_{j'}+C_{ii'}n_{j}n_{j'}+4n_in_jn_{i'}n_{j'}]\nonumber\\
%&&\!\! +w_3^{(k)}(\ell)[C_{ii'}C_{jj'}+C_{ji'}C_{ij'}
%-2n_ig_{i'j'}n_j-2n_{i'}g_{ij}n_{j'}+6n_in_jn_{i'}n_{j'}].
%\label{bitensor}
%\end{eqnarray}
\begin{equation}
 W_{(k)}{}^{ab}{}_{a'b'}(\ell)=w_k^{(1)}(\ell)t^{(1)}{}^{ab}{}_{a'b'}
+w_k^{(2)}(\ell)t^{(2)}{}^{ab}{}_{a'b'}
+w_k^{(3)}(\ell)t^{(3)}{}^{ab}{}_{a'b'}
\label{bitensor}
\end{equation}
where the tensors $t^{(I)}{}^{ab}{}_{a'b'}$ ($I=1,2,3$) are independent
of $k$, and are given by
\begin{eqnarray}
 t^{(1)}{}^{ab}{}_{a'b'}&=&(g^{ab}-3n^an^b)(g_{a'b'}-3n_{a'}n_{b'}),
\nonumber\\
t^{(2)}{}^{ab}{}_{a'b'}&=&C^{b}{}_{b'}n^an_{a'}+C^{a}{}_{b'}n^bn_{a'}
+C^{b}{}_{a'}n^an_{b'}+C^{a}{}_{a'}n^{b}n_{b'}+4n^an^bn_{a'}n_{b'},\nonumber\\
t^{(3)}{}^{ab}{}_{a'b'}&=&C^{a}{}_{a'}C^{b}{}_{b'}+C^{b}{}_{a'}C^{a}{}_{b'}
-2n^ag^{a'b'}n^b-2n_{a'}g^{ab}n_{b'}+6n^an^bn_{a'}n_{b'}.
\label{t123}
\end{eqnarray}
This is the general form which is traceless. It is known that the
bitensor depends only on the following quantities~\cite{allen,
allenjacobson}: geodesic distance $\ell$, unit vectors $n_{a}$,
$n_{a'}$, which are tangent to the geodesic at the point $x$ and
$x'$ (vectors or tensors with unprimed (primed) indices are the ones
defined at the point $x$ ($x'$); $n_a$ and $n_{a'}$ are defined to
be pointing away from the other point), metrics at each point, and
an operator $C_{ab'}$ which parallel transports a vector from one
point to the other along the geodesic.
%The expression (\ref{bitensor}) is the general form which
%is traceless.
%where $n_i$ is a unit vector tangent to the geodesic and $g_{ij}$
%is the linear map that parallel transports a vector along the
%geodesic.

The coefficient functions $w^{(I)}_k$
%$w_{1}^{(k)}(\ell)$, $w_{2}^{(k)}(\ell)$, $w_{3}^{(k)}(\ell)$,
depend only on the geodesic distance.
They are determined by the transversality condition and the
eigenvalue equation for the Laplacian on the hyperboloid
$\nabla^2_H W^{ab}_{(k)a'b'}=
-(k^2+3)W^{ab}_{(k)a'b'}$:
\begin{eqnarray}
w_k^{(1)}(\ell)&=&\frac{\sinh^{-5}\ell}{4\pi^2(k^2+1)}\Bigg\{\frac{\sin
k\ell}{k}\left(3+(k^2+4)\sinh^2\ell-k^2(k^2+1)\sinh^4\ell\right)\nonumber\\
&&~~~~-\cos k\ell\left(3/2+(k^2+1)\sinh^2\ell\right)\sinh 2\ell\Bigg\}\nonumber\\
w_k^{(2)}(\ell)&=&\frac{\sinh^{-5}\ell}{4\pi^2(k^2+1)}\Bigg\{\frac{\sin
k\ell}{k}\left(3+12\cosh \ell-3k^2(1+2\cosh \ell)\sinh^2\ell+k^2(k^2+1)\sinh^4\ell\right)\nonumber\\
&&~~~~+\cos k\ell\left(-12-3\cosh \ell+2(k^2-2)\sinh^2\ell+2(k^2+1)\cosh \ell\sinh^2\ell\right)\sinh \ell\Bigg\}\nonumber\\
w_k^{(3)}(\ell)&=&\frac{\sinh^{-5}\ell}{4\pi^2(k^2+1)}\Bigg\{\frac{\sin
k\ell}{k}\left(3-3k^2\sinh^2\ell+k^2(k^2+1)\sinh^4\ell\right)\nonumber\\
&&~~~~+\cos k\ell\left(-3/2+(k^2+1)\sinh^2\ell\right)\sinh 2\ell\Bigg\}
\label{w}
\end{eqnarray}

\subsubsection{Behavior near the boundary}
Let us write the bitensor explicitly when the two points are near
the boundary (spatial infinity).
%Let us now write the graviton correlator explicitly.
%using the Now let us see how the correlators behave when
%the two points approach the boundary.
We will use the Poincar\'{e} coordinate on the hyperboloid
\begin{equation}
 ds^2={1\over z^2}(dz^2+dx^2+dy^2),
\end{equation}
where the boundary is at $z=0$, and is parametrized
by $x$ and $y$. We consider two points
at $(x,y,z)$ and $(-x,-y,z)$; we will take the
$z\to 0$ limit eventually.

We can obtain the geometric quantities that appear in
the bitensor by recalling that the geodesic which connects two
points on the boundary is a half circle in the
Poincare coordinate.
The unit tangent vector $n_a$ at $(x,y,z)$ is
\begin{equation}
\pmatrix{n_z\cr n_x\cr n_y}={\sqrt{x^2+y^2}\over z\sqrt{x^2+y^2+z^2}}
\pmatrix{-1 \cr {x\over x^2+y^2}z\cr {y\over x^2+y^2}z},
%\quad\pmatrix{\tilde{n}_z\cr \tilde{n}_x\cr \tilde{n}_y}
%={\sqrt{x^2+y^2}\over z\sqrt{x^2+y^2+z^2}}
%\pmatrix{-1 \cr -{x\over x^2+y^2}z\cr -{y\over x^2+y^2}z},
\label{na}
\end{equation}
and $n_{a'}$ at $(-x,-y,z)$ is given by replacing $x$ and $y$
with $-x$ and $-y$ in the above equation\footnote{Both
$n_a$ and $n_{a'}$ are pointing in the direction which
increases the geodesic distance.}.

The parallel transport operator $C_{ab'}$ transforms
the tangent vector at one point to the one at the other point,
$n_{a}=-C_{a}{}^{b'}n_{b'}$, and it
preserves the norm of a vector.
In fact, in the present case where the $z$ coordinate of
the two points are equal, $C_{ab'}$ is a rotation matrix,
and we find
\begin{equation}
\pmatrix{
C_z{}^z & C_z{}^x & C_z{}^y\cr
C_x{}^z & C_x{}^x & C_x{}^y\cr
C_y{}^z & C_y{}^x & C_y{}^y\cr}
={1\over x^2+y^2+z^2}\pmatrix{
-x^2-y^2+z^2 & -2zx & -2zy \cr
2zx & -x^2+y^2+z^2 & -2xy \cr
2zy & -2xy & x^2-y^2+z^2}
\label{cab}
\end{equation}

Geodesic distance between $(x,y,z)$ and $(-x,-y,z)$ is
\begin{equation}
 \ell= \int  ds \sqrt{g_{ab}{d x^a\over
ds}{d x^b\over ds}}
=2 \log {\sqrt{x^2+y^2+z^2}+\sqrt{x^2+y^2}\over z}.
\end{equation}

We substitute these expressions into the bitensor and
obtain the correlator (\ref{gravlorentz}) in an expansion in $z$.

%We now study the asymptotic behavior of (\ref{gexpand})
%when the two points approach the boundary.
%We will be interested mostly in the piece which
%depends on $\bar{T}$. As in the
%scalar case, we also take the late time limit.
%We ignore terms with positive powers of $e^{-(\bar{T}+\ell)}$,
%and organize the remaining piece
% in an expansion in powers of $e^{\bar{T}-\ell}$.

%As we see from eqs. (\ref{gxasympt})-(\ref{w}),
%the integrand contains a piece which involve
%a factor $e^{-ik(\bar{T}-\ell)}$ and a piece which involve
%$e^{-ik(\bar{T}+\ell)}$. For the former, we close the
%contour in the upper half plane, and get terms
%with factors $e^{n(\bar{T}-\ell)}$ with $n=2,3,\ldots$,
%from the residues of the poles. For the latter, we close
%the contour in the lower half plane. We pick up a double
%pole at $k=i$. The leading pieces that are important in our
%discussion come from this pole. We get terms with positive
%powers of $e^{-(\bar{T}+\ell)}$ from the single poles in the
%lower half plane, which we can neglect in the
%$\bar{T}+\ell\to \infty$ limit.

\subsection{Large distance behavior of the correlator}
We study the large distance, late time behavior of the correlator
(\ref{gravlorentz}) in exactly the same way as we did for the scalar
in section A.2.3.
We will concentrate on the piece which depend on $\bar{T}$ (the
second term of (\ref{gravlorentz})).

First, note that the bitensor (\ref{w})
consists of pieces proportional
to $e^{-ik\ell}$ and those proportional to $e^{ik\ell}$;
the coefficient functions $w^{(I)}_k$ are
of the form $w^{(I)}_k=\hat{w}_k^{(I)}e^{-ik\ell}
+\tilde{w}_k^{(I)}e^{ik\ell}$, and $\hat{w}^{(I)}_k$ and
$\tilde{w}_k^{(I)}$ goes like $e^{-\ell}$ in the large $\ell$ limit.
Also note that the tensor structures
$t^{(I)}{}^{ab}{}_{a'b'}$ starts from order 1 terms.

As in the scalar case, the integration for the terms involving
$e^{-ik(\bar{T}-\ell)}$ is performed along the contour in the
Figure~\ref{fig-Lorentz}~(a) which surrounds the single poles at
$k=3i,4i,\ldots$. (No pole at $k=2i$ for graviton case.) For the
terms involving $e^{ik(\bar{T}+\ell)}$, the integration is done
along the contour in Figure~\ref{fig-Lorentz}~(b); it surrounds the
double pole at $k=i$, and single poles at $k=0$ and in the lower
half plane. The  contribution from the $k=i$ double pole in the
latter integral yields the terms that we are interested in. The rest
of the terms are subleading for the reason we explained in section
A.2.3.
%contour in the upper half plane for the terms involving
%$e^{-ik(\bar{T}-R)}$, and in the lower half plane for the terms
%involving $e^{ik(\bar{T}+R)}$. (See Figure.) The latter integral
%picks up a contribution from the $k=i$ double pole, and this
%gives all the terms that we are interested in. Contributions
%from the other poles are subleading for the same reason
%as in the scalar case.

The contribution from the $k=i$ double pole takes the form
\begin{equation}
 \langle h^{ab}h_{a'b'}\rangle_{(k=i)}
=ce^{-\bar{T}}\sum_{I=1}^{3}
\partial_k\left((k-i)\rc (k){e^{-ik\bar{T}}
\over k}\hat{w}^{(I)}_k(\ell)e^{-ik\ell}\right)
\bigg|_{k=i}t^{(I)}{}^{ab}{}_{a'b'}
\label{gravdoublep}
\end{equation}
where $c$ is a constant, and the $e^{-\bar{T}}$ factor is due to the
rescaling to get the physical correlator.
%We have defined
%$\hat{w}^{(I)}_k$ ($I=1,2,3$) to be the coefficient
%of $e^{-ik\ell}$ in $w^{(I)}_k$,
%$w^{(I)}_k=\hat{w}_k^{(I)}e^{-ik\ell}
%+\tilde{w}_k^{(I)}e^{ik\ell}$.

\subsubsection{Dimension 0 piece}

The leading term of the correlator grows linearly
with the geodesic distance $\ell$. This term arises
when the $k$-derivative in (\ref{gravdoublep})
hits $e^{-ik\ell}$.

%As in the scalar case, the correlator grows linearly with
%the geodesic distance $\ell$.
%The leading term of $\hat{w}^{(I)}_{k=i}$ goes
%like $e^{-\ell}$, which cancels the factor $e^{-ik\ell}$.
%We have a linear term in $\ell$ when the derivative hits
%$e^{-ik\ell}$.

The tensor structure of this piece is given by the leading
term of $\sum_I \hat{w}_{(k=i)}^{(I)}t^{(I)ab}{}_{a'b'}$.
Substituting (\ref{w}) into $\hat{w}_{(k=i)}^{(I)}$, and
(\ref{t123}), (\ref{na}), (\ref{cab}) into $t^{(I)ab}{}_{a'b'}$,
we find that the tensor structure of the dimension 0 piece
is as follows.
%The tensor structure of this term is found by
%taking the leading term of
%$\sum_I \hat{w}_{(k=i)}^{(I)}t^{(I)ab}{}_{a'b'}$.
%Let us see its tensor structure by taking
%the $\ell\to\infty$ ($z\to 0$) limit in
%$w_k^{(I)}$ and $t^{(I)}{}^{ab}{}_{a'b'}$ given by
%(\ref{w}) and (\ref{t123}).
If the correlator has at least one $z$ index,
it is of order $O(z^2)$; in other words,
the order 1 correlator has all its indices
along the boundary directions,
\begin{equation}
 \langle h^{ij}h_{i'j'}\rangle_{(0)}
=c_{0}\ell\, t{}^{ij}{}_{i'j'}
\label{gravitylog}
\end{equation}
where $c_{0}$ is a constant, and
$i,j=1,2$ denotes the indices along the 2 dimensional
plane along the boundary.
The tensor $t{}^{ij}{}_{i'j'}$ is
%independent of $z$. It is
traceless in the 2-D sense, $t{}^{i}{}_{ii'j'}=0$,
since the correlator
is traceless on the hyperboloid by construction,
and the radial component is zero to this order.
%
%It is given by
%\begin{equation}
% t^{ij}{}_{i'j'}=g^{ij}g_{i'j'}-\hat{C}^{i}{}_{i'}\hat{C}^{j}_{j'}
%-\hat{C}^{i}{}_{j'}\hat{C}^{j}{}_{i'},
%\end{equation}
%where $\hat{C}^{i}{}_{j'}$ is the leading piece ($z\to 0$ limit)
%of the 2-D part of the matrix (\ref{cab}).
%
Explicit form when the two points are at $(x,y,z)$ and
$(-x,-y,z)$ is
\begin{equation}
 t_{x}{}^{x}{}_{x}{}^{x}=-t_{x}{}^{x}{}_{y}{}^{y}
=-t_{x}{}^{y}{}_{x}{}^{y}=
-{(x^4-6x^2y^2+y^4)\over (x^2+y^2)^2},\quad
t_{x}{}^{x}{}_{x}{}^{y}=-t_{y}{}^{y}{}_{x}{}^{y}
=-{4(x^2-y^2)xy \over (x^2+y^2)^2}.
\label{tensort}
\end{equation}
The geodesic distance is given by $\ell=2\log {2\sqrt{x^2+y^2}\over z}$
to the leading order.
%Other components are given by the traceless condition in 2-D.
%(which follows from the fact that the tensor is traceless in
%hyperboloid, and the $z$-$z$ components are zero to this order).

Let us calculate a gauge invariant quantity associated with
this correlator. We find that the 2-D scalar curvature
of the graviton fluctuation has finite ($z$-independent)
correlation when the two points approach the boundary.
The scalar curvature is given by
$C=h_{i}{}^{i}{}_{;j}{}^{;j}
-h_{i}{}^{j}{}_{;j}{}^{;i}-R_{i}{}^{k}h_{j}{}^{i}$,
where the semi-colon denotes the covariant derivative
in the background 2-D space, and $R_{i}{}^{j}$ is the 2-D
Ricci tensor of the background.
The first and the third terms
are zero, since the fluctuation
is traceless and the background is flat.
From (\ref{gravitylog}), we get
\begin{equation}
 \langle C(x_1)C(x_2)\rangle =
\partial^{i}\partial_{j}\partial^{i'}\partial_{j'}
\langle h_{i}{}^{j}(x_1) h_{i'}{}^{j'}(x_2)
\rangle_{(0)}
={2c_{0}\over \left\{(x^{i}_1-x^{i}_2)(x^{i}_1-x^{i}_2)\right\}^2}.
\end{equation}
%There is no logarithmic behavior in this correlator.
The 2-D curvature behaves as a dimension 2 field.
%In the large $\ell$ limit, $\hat{w}^{(k=i)}_I$ goes like $e^{-\ell}$,
%as we see from (\ref{w}). The correlator has a piece which
%grows linearly with $\ell$, which arises when the derivative
%hits $e^{-ik\ell}$.

\subsubsection{Transverse-traceless (TT) dimension 2 piece}
Let us now study the subleading terms. Let us first
look for the terms
which do not contain the linear factor in $\ell$.

We ignore the pure gauge piece which does not contribute
to the physical quantities, as we have ignored the constant
term in the scalar case. Note that the bitensor
$W_{(k=i)}^{ab}{}_{a'b'}$ evaluated at $k=i$
is a pure gauge. It is known that the tensor harmonics on $\hyp$
at $k=i$, which satisfy $\nabla^{2}_{H}q_{ab}=-2q_{ab}$, can be
written in the form of a gauge transformation $q_{ab}=\nabla_{a}v_{b}
+\nabla_{b}v_{a}$
with a gauge parameter $v_a$ satisfying $\nabla^2_{H}
v_a=2v_a$~\cite{sasaki,turok2}.
%($q_{ab}$ and $v_a$ do not decay at infinity.)
Thus, we can add a term proportional (by an $\ell$-independent factor)
$W_{(k=i)}^{ab}{}_{a'b'}=\sum_I(\hat{w}^{(I)}_{k=i}
e^{\ell}+\tilde{w}^{(I)}_{k=i}e^{-\ell})t^{(I)}{}^{ab}{}_{a'b'}$
to the correlator without affecting the physical
properties.

%In the literature~\cite{sasaki,turok2}, it seems that the whole
%contribution from the $k=i$ pole is considered as pure gauge and
%ignored, contrary to our discussion here. We are finding important
%physical pieces form the $k=i$ pole.

In the literature~\cite{sasaki,turok2}, it is stated that the
``supercurvature mode'' (non-normalizable mode) is absent for tensor
perturbations. It seems that the whole contribution from the $k=i$
pole is neglected assuming it is pure gauge, contrary to our
discussion here. If we start from the Euclidean prescription, we
find no reason to neglect the $k=i$ pole contributions except for
the pure gauge term mentioned above. We are finding important
physical pieces from this pole.

%contribution from the $k=i$ pole is considered as pure gauge and
%ignored, contrary to our discussion here. We are finding important
%physical pieces form the $k=i$ pole.

In the double pole contribution (\ref{gravdoublep}), there
is a term where the $k$-derivative hits $(k-i)\rc (k)/k$. This
term is proportional to $\sum_I\hat{w}^{(I)}_{k=i}e^{\ell}t^{(I)}
{}^{ab}{}_{a'b'}$ and is of order 1. Up to pure gauge, this
term can be written in the form $\sum_I\tilde{w}^{(I)}_{k=i}
e^{-\ell}t^{(I)}{}^{ab}{}_{a'b'}$, which is of order $z^4$
(which corresponds to dimension 2),
%This term has the same tensor structure
%as the previous piece,
%By expanding $w_k^{(I)}$ and $t^{(I)}{}^{ab}{}_{a'b'}$ in $z$,
%we find that the next to leading term of the correlator,
%besides the terms containing logarithm and the constant term, is
%Besides the terms containing the logarithm and the constant
%term, the next to leading term of the $\langle h^{a}_{b}
%h^{a'}_{b'}\rangle$ is the dimension 2 piece. By
%This has the same tensor structure as the linearly growing
%term, $t^{ij}{}_{i'j'}$ given by (\ref{tensort})
\begin{equation}
 \langle h^{ij}h_{i'j'}\rangle_{(2)}=
{c_{2}z^4\over (x^2+y^2)^2}t{}^{ij}{}_{i'j'},
\end{equation}
where $c_2$ is a constant and the tensor
$t^{ij}{}_{i'j'}$ is given by (\ref{tensort}).
The correlator $\langle h^{ij}h_{i'j'}\rangle_{(2)}$ is traceless
in 2-D, and is also transverse in the 2-D space,
\begin{equation}
 \partial_i  \langle h^{ij}h_{i'j'}\rangle_{(2)}=0,
\end{equation}
as we can check explicitly. Or more simply, we can
write the correlator in the complex
coordinate $u=x+iy$, and check that it is holomorphic
$\langle h^{uu}h^{uu}\rangle \sim 1/u^4$,
$\langle h^{uu}h^{\bar{u}\bar{u}}\rangle =0$.

\subsubsection{Non-TT dimension 2 piece}

In the $k=i$ contribution (\ref{gravdoublep}), there is one more
piece, which arises when the $k$-derivative hits the coefficient
function $\hat{w}^{(I)}_k$. By subtracting the pure gauge, we can
show that the leading term from this piece is of order $z^4$
(dimension 2). However, this dimension 2 piece has a tensor
structure different from the above one, and is neither transverse
nor traceless in the 2-D sense.

Also, as in the scalar case, the dimension 0 (logarithmic) term
has subleading corrections and gives rise to the dimension 2 piece
with the logarithmic prefactor. This piece is also non-TT in 2-D.

%These terms are similar to the dimension 0 piece in the sense that
%the structure of the bitensor is broken (since
%$k$-derivative hits the bitensor). This might suggest that
%these pieces are descendant of the dimension 0 field, but as we
%mention in the text, we have no clear interpretation of these
%terms in the CFT.

To interpret these terms, let us temporarily assume that the pole of
$\rc (k)$ is at a slightly shifted position $k=(1-\epsilon)i$. As we
have discussed in section A.4 for the scalar case, we will have
separate contributions from the single poles at $k=i$ and
$k=(1-\epsilon)i$. The former is proportional to
$\tilde{w}_ke^{ik\ell} t^{(I)}{}^{ab}{}_{a'b'}|_{k=i}$. This always
has dimension 2 (after subtracting the pure gauge term), and is TT
in 2-D. In the text, we identify this piece as a candidate for the
energy-momentum tensor of the CFT. The contribution from the
$k=(1-\epsilon)i$ pole is proportional to $\hat{w}_ke^{-ik\ell}
t^{(I)}{}^{ab}{}_{a'b'}|_{k=i}$. This term yields pieces with
dimension $\epsilon$ and $2+\epsilon$, which are non-TT in 2-D.

The non-TT dimension 2 piece in the massless graviton correlator
that we found above is a consequence of merging of the poles. As in
the scalar case, the dimension 2 piece and dimension $2+\epsilon$
piece have opposite signs, and in the massless limit $\epsilon \to
0$, they give rise to terms which involve the derivative of the
bitensors with respect to $k$.

\end{document}